\documentclass[twocolumn]{aastex631}

\usepackage{amsmath,amstext}
\usepackage{gensymb}
\usepackage{subfigure}
\usepackage[T1]{fontenc}
\usepackage{xcolor}
\usepackage{stackengine}\setstackEOL{\cr}

\usepackage{appendix}

\usepackage{longtable}

\begin{document}
\title{Mass Segregation in the CMZoom Survey}
\author{\href{https://orcid.org/0009-0008-8140-7334}{S F Schuler}}\footnote{Email: stefania.schuler@uconn.edu}
\affiliation{University of Connecticut, Department of Physics, 196A Auditorium Road, Unit 3046,
Storrs, CT 06269}

\author{\href{https://orcid.org/0009-0002-7459-4174}{J Wallace}}
\affiliation{University of Connecticut, Department of Physics, 196A Auditorium Road, Unit 3046,
Storrs, CT 06269}

\author{\href{https://orcid.org/0000-0002-6073-9320}{C Battersby}}
\affiliation{University of Connecticut, Department of Physics, 196A Auditorium Road, Unit 3046,
Storrs, CT 06269}

\author{\href{https://orcid.org/0000-0003-0946-4365}{H P Hatchfield}}
\affiliation{University of Connecticut, Department of Physics, 196A Auditorium Road, Unit 3046,
Storrs, CT 06269}

\author{\href{https://orcid.org/0000-0002-6447-899X}{R Gutermuth}}
\affiliation{University of Massachusetts,
Department of Astronomy, LGRT 849, 710 North Pleasant Street, Amherst, MA 01003-9305}

\author{\href{https://orcid.org/0000-0003-2619-9305}{X Lu}}
\affiliation{Shanghai Astronomical Observatory, Chinese Academy of Sciences, 80 Nandan Road, Shanghai 200030, P.R. China}

\author{\href{https://orcid.org/0000-0002-8389-6695}{S Zhang}}
\affiliation{Shanghai Astronomical Observatory, Chinese Academy of Sciences, 80 Nandan Road, Shanghai 200030, P.R. China}

\author{\href{https://orcid.org/0000-0003-2384-6589}{Q Zhang}}
\affiliation{Harvard-Smithsonian Center for Astrophysics, 60 Garden Street, Cambridge, 02138, USA}

\begin{abstract}

We employ a Minimum Spanning Tree (MST) approach to characterize the spatial distribution and mass segregation of compact millimeter continuum sources within the Central Molecular Zone (CMZ) of the Milky Way. We use a modified form of the complete version of the 1.3 mm dust continuum catalog from the CMZoom survey, which identifies 685 compact sources with typical effective radii of $\sim0.1$ pc. For 22 of 35 CMZ clouds, we calculate the thermal and turbulent Jeans lengths and masses, and determine that compact source separations, as well as compact source masses, are more consistent with thermal fragmentation at $\sim0.1$ pc size scales. We construct the mass segregation ratios for compact sources in 17 CMZ clouds and determine that 5 of the analyzed clouds display some form of mass segregation ($\Lambda_{MSR} > 1.5$), while the remaining clouds show either inverse mass segregation ($\Lambda_{MSR} < 0.75$), or no evidence of true mass segregation ($0.75 < \Lambda_{MSR} < 1.5$). Finally, we find that although some actively star-forming clouds do exhibit mass segregation, other similarly active clouds do not, indicating an unclear correlation with evolutionary stage for star forming clouds in the CMZ, given the current available data.

\end{abstract}


\section{Introduction} \label{sec:intro}

The Central Molecular Zone (CMZ) consists of the inner few hundred pc of the Galactic Center \citep{Henshaw_2022, Battersby_2025a}, and contains roughly 5\% of the molecular gas in the Milky Way \citep{Dahmen_1998}, most of which is concentrated in the numerous dense molecular clouds such as G0.699-0.028 (Sgr B2), G359.484-0.132 (Sgr C), and the Dust Ridge \citep{Mills_2018}. It is an extreme environment, with a high gas number density of $n_{\text{H}_2}$ $>$ $10^{4}$ cm$^{-3}$ (e.g., \citealt{Paglione_1998}; \citealt{Jones_2013}), extreme temperatures $>$ 60 K in comparison to the Galactic disk \citep{Ginsburg_2016}, and magnetic field strengths of $\sim 1$ mG for some clouds \citep{Mangilli_2019, Federrath_2016}. \cite{Lu_2024} find magnetic field strengths in several molecular clouds in the CMZ to range between 0.1 mG to 1.73 mG.

Despite the abundance of dense gas, the CMZ is widely considered to be under-producing stars overall \citep{Kauffmann_2017a, Immer_2012, Guesten_1983, Beuther_2012, Barnes_2017}. Using the \textit{CMZoom} survey, \cite{Hatchfield_2020} put a lower and upper limit on the potential for star-formation in the CMZ of $\approx$ 0.04 to 0.47 $M_{\odot} yr^{-1}$ (with the exclusion of G0.699-0.028; $\approx$ 0.08 to 2.20 $M_{\odot} yr^{-1}$ inclusive of G0.699-0.028). \cite{Hatchfield_2024} argue that the incipient star formation rates (SFRs) in the CMZ might in fact be higher than previously estimated, due to the fact that use of the cold dust continuum alone cannot fully differentiate between active star-formation and quiescent dust, and calculate limits similar to those found by \cite{Hatchfield_2020}. 

We study mass segregation within star-forming clouds in the CMZ as selected by the \textit{CMZoom} Survey \citep{Battersby_2020, Hatchfield_2020}. Prior studies have investigated the mass segregation of stellar clusters to impose constraints on their formation and evolution \citep{Pouteau_2023, Pavl_k_2019, Dib_2019}. These studies have used a variety of methods such as the mass segregation ratio (MSR), which uses minimum spanning trees (MSTs) to characterize mass distribution in a cluster \citep{Allison_2009a}.
 
More recently, these analyses have been performed on clouds that are forming pre- and protostellar cores \citep{Kinman_2024, Morii_2023}.
\cite{Nony_2021} in particular, use \cite{Allison_2009a}'s method to calculate the mass segregation for clouds and concludes that initial mass segregation of gas in dense prestellar cores may contribute to the eventual mass segregation of stars in clusters. \cite{Sanhueza_2019} find no strong evidence of early (primordial) mass segregation of prestellar cores in the infrared dark clouds surveyed, but do conclude that future core accretion may result in dynamic core mass segregation. \cite{Dib_2019} examine (prestellar and dense) cores in four star-forming clouds and conclude that three of those clouds exhibit primordial mass segregation, where the cores initially form within the densest parts of the clouds. 

The current state of knowledge on the mass-distribution of compact sources in the CMZ is limited to a few well-researched clouds such as the G0.699-0.028 cloud, G359.484-0.132 cloud, and G0.253+0.016 (the Brick) \citep{Kinman_2024}. When examining millimeter CMZ sources on the scale of $\approx 1000$ au, \cite{Kinman_2024} found that G359.484-0.132 shows signs of mass segregation and star-formation, while G0.253+0.016 shows signs of neither, and make note that the most massive prestellar cores observed are expected to be located in the densest parts of the cloud, in order to facilitate future accretion.

In order to gain a more complete understanding of compact source mass segregation in the CMZ, we need to leverage
millimeter-wavelength observations of the dust continuum 
at sub-parsec scales that are sufficient to resolve the individual star-forming clouds throughout the entire CMZ. The \textit{CMZoom} survey is the first large-scale survey of the CMZ at arcsecond / sub-parsec resolution  \citep{Battersby_2020}. This makes it a good candidate for studying mass segregation throughout the densest star-forming clouds of the CMZ in a uniform manner. Future studies, such as the ALMA CMZ Exploration Survey, ACES, will help further these analyses \citep{Longmore_2026}. 

In this paper, we investigate the distribution and mass segregation of compact sources for a complete sample of star-forming clouds located in the Central Molecular Zone. In Section \ref{sec:data} we describe the \textit{CMZoom} Survey SMA observations and the \textit{CMZoom} compact source catalog. We describe our method for characterizing fragmentation throughout the CMZ in Section \ref{sec:tjl} and our MSR analysis using MSTs in Section \ref{sec:msts}. We report the results of our analyses in Section \ref{sec:results} and discuss the broader implications of our results in Section \ref{sec:discussion}.

\begin{figure*}[htb!]
\includegraphics[width=\textwidth]{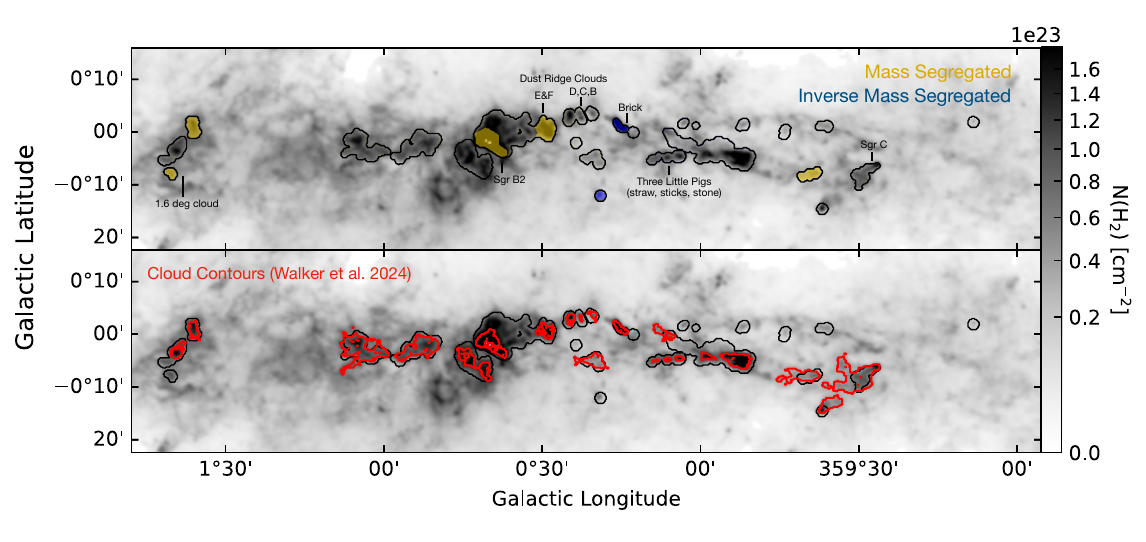}
\caption{An overview of the clouds covered in the CMZoom survey. The Herschel Hi-GAL column density map of the CMZ \citep{Battersby_2024a} is shown in gray scale, and black contours represent the area covered by the CMZoom Survey. The top panel describes each cloud by its level of mass segregation as determined from its Mass Segregation Ratio (MSR) in Section \ref{sec:msrs}. Regions that are shaded yellow are considered mass segregated, whereas the blue shaded regions are considered inverse mass segregated. Unshaded regions represent the clouds that do not have enough source detections to make a meaningful MSR calculation or can represent regions that do not meet the MSR thresholds for (inverse) mass segregation. A number of well-known CMZ clouds are labeled using their colloquial names.  The bottom panel shows the CMZ clouds defined in \cite[][red contours]{Walker_2024} that are associated with our CMZoom regions, which we use to measure thermal and turbulent Jeans properties in Section \ref{sec:tjl}.}
\label{fig.overview}
\end{figure*}

\section{Summary of Data}
\label{sec:data}
\subsection{SMA Observations}
\label{subsec:obs}
 The \textit{CMZoom} Survey is a large Submillimeter Array (SMA) program that observed all of the high-column-density gas (N(H$_{2}) > 10^{23}$ cm$^{-2}$) contained within the CMZ \citep{Battersby_2020, Hatchfield_2020}. The 1.3 mm continuum emission data have an angular resolution of $\sim$3\arcsec, corresponding to a typical spatial resolution of $\sim$0.1 pc at an assumed distance of 8.178 kpc \citep{GRAVITY_2019}. A total of 35 clouds were observed covering an area of $\sim$240 square arcminutes ($\sim$1360 pc$^{2}$). All but 4 of the 35 clouds (G1.670-0.130, 
G0.393-0.034, G0.212-0.001, and G359.137+0.031) are confirmed to be located in the CMZ by \cite{Hatchfield_2024}, and \cite{Battersby_2020} find that G0.316-0.201 and G359.615-0.243 are not in the CMZ. Additionally, the \textit{CMZoom} program observed spectral line emission from a variety of molecular species, including CO, H$_2$CO, and CH$_3$OH \citep{Callanan_2023}.

\subsection{CMZoom Compact Continuum Source Catalog}
\label{subsec:cat}
The analysis we perform in this paper makes use of a version of the complete \textit{CMZoom} Survey compact dust continuum source catalog, with an original total of 816 sources \citep{Hatchfield_2020}. Since the primary beam corrected SMA images have enhanced noise at the edges, we remove catalog sources within 0.5 pc of the map boundary, to reduce the impact of possible false detections on our mass segregation analysis. This reduced version of the complete CMZoom catalog has a total of 685 sources.  This catalog was generated using a pruned dendrogram-based technique, \verb|astrodendro|\footnote{Refer to \url{http://www.dendrograms.org/} for additional information on the astrodendro package, and to \cite{Hatchfield_2020} section 4.3, Appendices A and B for further information on the chosen parameters.}, which is a Python package that takes astronomical data and decomposes it into a dendrogram, or a hierarchical representation of the data. The dendrogram is a tree-like diagram where the densest structures that contain no additional substructures are categorized as ``leaves'', whereas structures containing additional substructure are known as ``branches'' and ``trunks''. The only structures included in the \textit{CMZoom} catalog are the densest ``leaf'' structures (or ``compact sources''), with typical effective radii between 0.04 and 0.4 pc at an assumed distance of 8.178 kpc.  

The dendrogram source extraction performed by \cite{Hatchfield_2020} required structures to have a minimum pixel value larger than 3$\sigma_{global}$ and a minimum leaf boundary significance of 1$\sigma_{global}$, where $\sigma_{global}$ is the minimum root-mean-squared (RMS) noise estimate for the entire surveyed cloud, 3 MJy sr$^{-1}$. Each structure contains a minimum number of pixels equivalent to half the typical beam size. Additionally, the cataloged leaf structures were required to meet local noise thresholds. The local RMS noise estimates were calculated using SMA residual maps, as detailed in \cite{Battersby_2020}. The complete version of the ``pruned'' CMZoom catalog only includes sources with a peak flux of at least $4\sigma_{local}$, while the robust catalog requires sources to have a peak flux of at least $6\sigma_{local}$. Both the complete and robust catalogs exclude leaves with a mean flux $<2\sigma_{local}$. While the complete catalog is sensitive to lower mass sources and most of the sources unique to the complete catalog are in the low-mass range and contain roughly 13\% of the mass of the robust catalog, there is a higher frequency of false positive detections, although \cite{Hatchfield_2020} finds that a significant number of false positives are unlikely in the lower-mass range. Conversely, the robust catalog prioritizes the reduction of false positive detections, but increases the chance of removing real, lower-mass sources. In this paper, we use the reduced form of the complete CMZoom catalog to ensure our analysis neither excludes low mass detections nor contains too many false positives. However, we also consider the effects of including only the most robust detections in our analysis, as discussed in Section \ref{sec:msrrob}. For a more detailed description of the applied dendrogram source extraction of the \textit{CMZoom} continuum data, we refer the reader to \cite{Hatchfield_2020}.

\subsection{CMZ Cloud Catalog}
\label{subsec:higal}

In Section \ref{sec:tjl}, we use the dust temperatures, gas densities, and gas velocity dispersions for molecular clouds in the CMZ to characterize their fragmentation properties. Our analysis makes use of the CMZ cloud catalog presented in \cite{Walker_2024} and \cite{Battersby_2025b}. 
 
This catalog was generated from the Herschel dust column density map from \cite{Battersby_2025a} using the Python package \verb|astrodendro| as described in \cite{Battersby_2025b}, with specific selection criteria applied in \cite{Walker_2024}. Further information about the \verb|astrodendro| input parameters and the precise implementation are provided in \cite{Walker_2024, Battersby_2025b}. The molecular clouds reported in this catalog have effective radii ranging from 1.4 pc -- 8.7 pc and masses ranging from $1 \times10^{4} - 1 \times 10^{6}$ M$_{\odot}$. The dust temperature map presented in 
\cite{Battersby_2025b} was used to estimate the median and peak temperature for each cloud. \cite{Walker_2024} derived cloud velocity dispersions by performing multiple component Gaussian fits on HNCO (4$_{0,4}$ - 3$_{0,3}$) and H$_{2}$CO (3$_{0,3}$ - 2$_{0,2}$) spectra\footnote{See Section 2 in \cite{Walker_2024} for information on the specific data.}. This is performed at angular resolutions ranging from 3\arcsec  to 39\arcsec and spatial resolutions $\ge 4$ pc, which are similar to or coarser than the \textit{CMZoom} survey resolution. Although both species are reliable tracers of dense gas in the CMZ, we use the velocity dispersions derived from the H$_{2}$CO spectral line fits for our analysis. We find that H$_{2}$CO has fewer components per cloud than HNCO in \cite{Walker_2024}, which simplifies the analysis. Nonetheless, the derived velocity dispersions for both H$_{2}$CO and HNCO are similar enough in magnitude that the use of either tracer does not change the conclusions of our results. 

\subsection{Overview of Clouds and Tables}

The \textit{CMZoom} Survey looks at multiple clouds which are areas of intense research in the CMZ, including G0.699-0.028, G359.484-0.132, G0.253+0.016, and the Dust Ridge regions (G0.489+0.010, G0.412+0.052, G0.380+0.050, and G0.340+0.055). We are unable to measure the mass segregation ratios for certain clouds (such as G359.484-0.132) which have too few sources in our modified version of the complete compact catalog. 

In this work, we use the non-background-subtracted masses for each source. Although this may be an overestimate of the true mass, we find that the choice in using the non-background-subtracted masses over the background-subtracted masses has a minimal impact on the classification of our clouds. The clouds that show the most significant change in MSR when using the background-subtracted sources instead of the non-background-subtracted sources can be seen in Appendix \ref{bgsmsr}, and are: G0.380+0.050 (Dust Ridge cloud C), G0.326-0.085, G0.316-0.201, and G359.889-0.093 (20 km/s cloud).

Table \ref{tab:mass} contains the calculated values for both the mass segregation analysis and the Jeans lengths and masses. Table \ref{tab:sfr} compares whether specific clouds display both mass segregation and detected star formation tracers (from \cite{Hatchfield_2024} and \cite{Battersby_2025b}, the latter of which uses the former's estimates).


\section{Thermal and Turbulent Jeans Length and Mass}\label{sec:tjl}

\subsection{Minimum Spanning Tree Method}\label{msts}

In order to investigate the type of fragmentation in the clouds included in this survey, we compare the mean edge length ($\bar{\ell}$) of the particular cloud's minimum spanning tree (MST) to both the cloud's thermal Jeans length, $\lambda_{th}$ and its turbulent Jeans length, $\lambda_{turb}$. 

An MST is the shortest path between all points in a group, where the individual paths between each point are known as ``edge lengths'' \citep{Prim_1957,Barrow_1985}. The MST uniquely allows for a view of the spatial separation between the compact sources without relying on such other methods as mean angular separation.  

\begin{figure}[h!]
\includegraphics[width=0.48\textwidth]{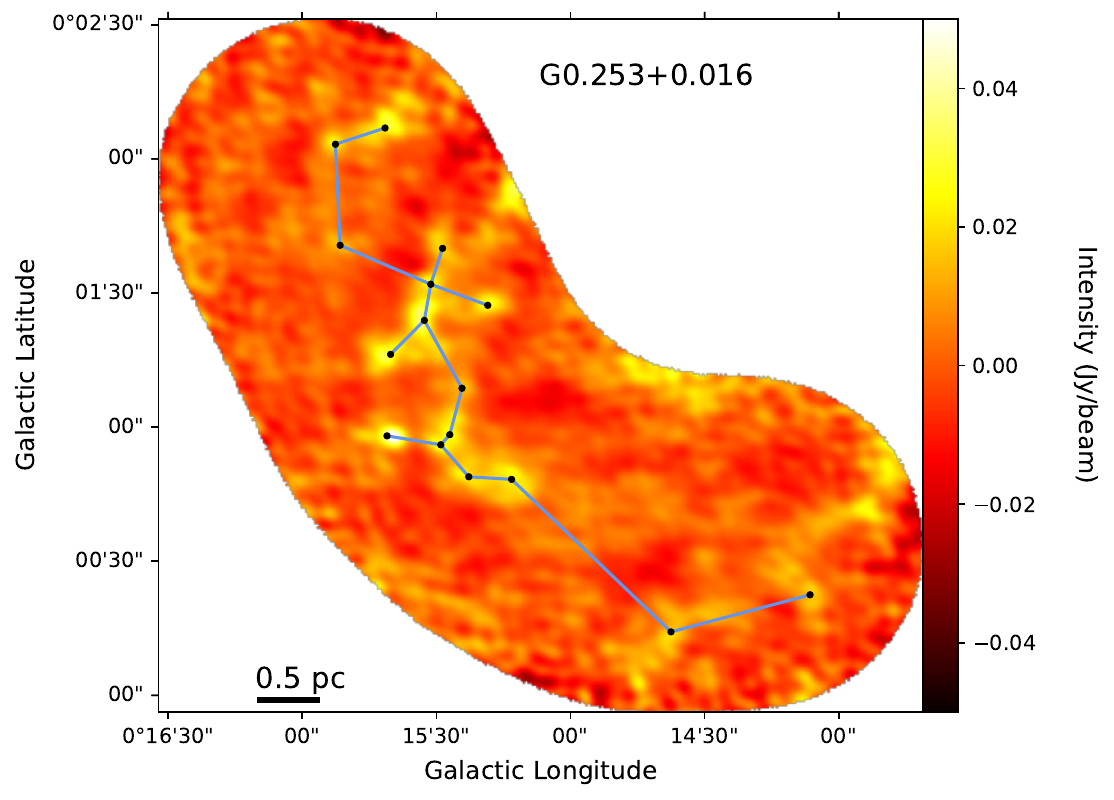}
\caption{Applying the MST algorithm as discussed in Section \ref{msts} to the case of the G0.253+0.016 cloud (shown), a simple elongated path. The black points represent the locations of the compact sources
based on their galactic coordinates. The blue lines connecting the black points are the edge lengths of the MST. The
1mm dust continuum SMA observation for the cloud is underlaid. The complete figure set (35 images) is available in the online journal.}
\label{fig.figuresetex1}
\end{figure}

In order to generate the MST for a cloud with more than two defined compact sources, we first build a sparse matrix composed of the angular separation (in degrees) between each combination of compact sources within a \textit{CMZoom} cloud. When this sparse matrix has been built for the cloud, we apply the scipy \verb|minimum_spanning_tree| package, which optimizes the shortest overall path using the distances in the matrix (see Figure \ref{fig.figuresetex1}). 

The presence of distant outlier detections may affect the construction of the MST, such as its shape and total length, which ultimately affects our analysis of these regions. We discuss how these outlier detections may affect our subsequent results in Section \ref{sec:err}.

\figsetstart
\figsetnum{1}
\figsettitle{\textit{CMZoom} Overview Figures}
\figsetgrpstart
\figsetgrpnum{1.17}
\figsetplot{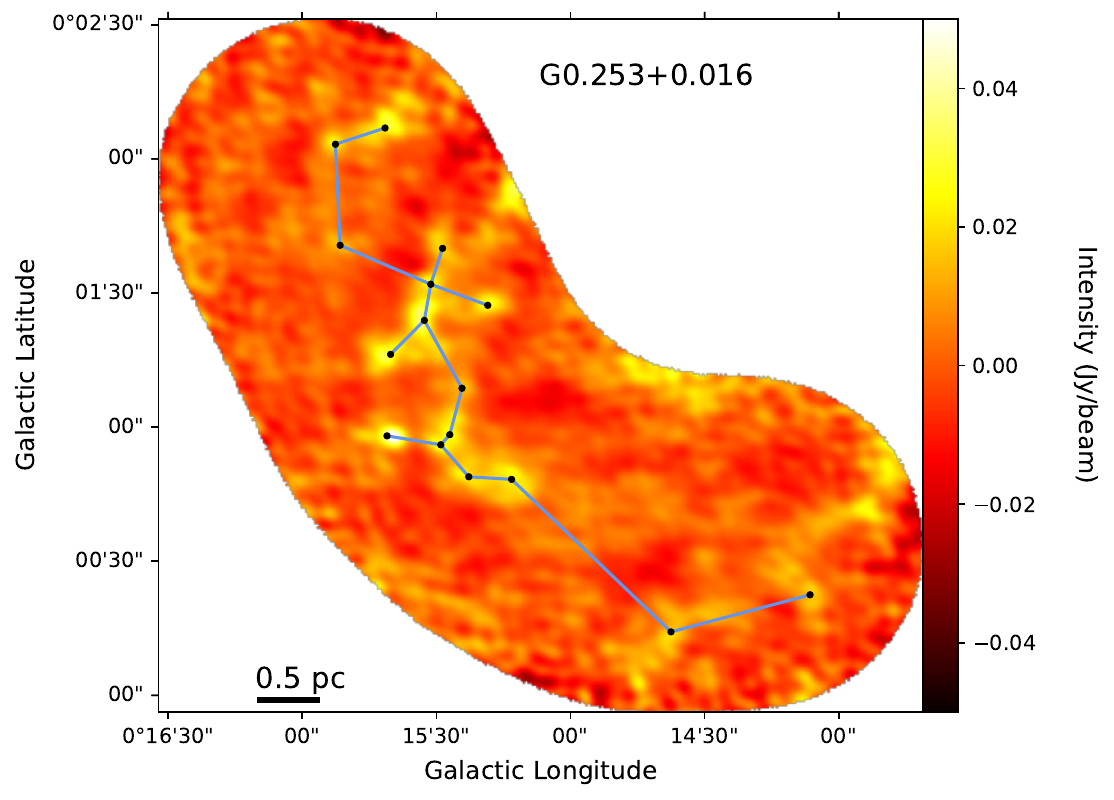}
\figsetgrpend
\figsetend

\subsection{Thermal Jeans Length and Mass}
\label{sec:tjlm}

We follow \cite{Sanhueza_2019}'s correction factor (division by $\frac{2}{\pi}$) to calculate the deprojected mean edge length ($\bar{\ell}_{\text{corr}}$), which we compare with $\lambda_{th}$ and  $\lambda_{\text{turb}}$. We report $\bar{\ell}_{\text{corr}}$ for each cloud in Column 8 of Table \ref{tab:mass}.

In this work, we follow the analysis of \cite{Zhang_2021} with the equation for $\lambda_{th}$ as follows, derived in \cite{Mac_Low_2004}. 

\begin{equation}
    \lambda_{th} = \sigma_{th}\sqrt{\frac{\pi}{G\rho}}
\end{equation} where 
\begin{equation}
    \sigma_{th} = \sqrt{\frac{k_{B}T}{\mu m_{H}}}
\end{equation}

We use a value of 2.37 for $\mu$ (the mean molecular weight per free particle; \citealt{Kauffmann_2008}), Herschel Hi-GAL median dust temperatures (ranging between 20 and 25 K) for each cloud, and Herschel Hi-GAL densities as $\rho$. 

We have also considered the thermal Jeans mass, $M_{th}$ \citep{Mac_Low_2004} as follows: 

\begin{equation}
    M_{th} = \frac{4\pi\rho}{3}\left(\frac{\lambda_{th}}{2}\right)^{3} = \frac{\pi^{5/2}}{6}\frac{\sigma_{th}^{3}}{\sqrt{G^{3}\rho}}
\end{equation}

In our calculations of the Jeans lengths and masses, we consider the main sources of error to be the dust temperature (a statistical error of roughly 10\%, please see Section \ref{sec:err} for more about the systematic uncertainty in the core temperatures), and the densities from the Herschel Hi-GAL table (an uncertainty of a factor of 2) \citep{Battersby_2025a}. These uncertainties have been included in the calculations of the Jeans lengths and masses, and are shown in Table \ref{tab:mass}. 


\subsection{Turbulent Jeans Length and Mass}

The turbulent Jeans length, $\lambda_{turb}$, and turbulent Jeans mass, $M_{turb}$, are calculated in the same way as $\lambda_{th}$ and $M_{th}$, but with non-thermal linewidths rather than sound speed (eg, in \citealt{Wang_2014}). We calculate $\lambda_{turb}$ and $M_{turb}$ for every cloud, using the Gaussian $\sigma$ widths converted from non-thermal FWHM linewidths for H$_2$CO from \cite{Walker_2024}, except for G0.316-0.201, which has no detected H$_2$CO lines.

The H2CO linewidths from \cite{Walker_2024} are derived at scales at or larger than that of the \textit{CMZoom} survey. The derivation at this scale potentially overestimates the velocity dispersions of the gas enclosing the star-forming cores, given that velocity dispersions have been shown to decrease at smaller physical scales \citep{Kauffmann_2017b}. We therefore consider a ``clump-scale'', which is defined as the smallest ellipse that encloses all sources within a region. The ellipses are applied to the column density and temperature maps produced by \cite{Battersby_2025b}; the column densities within the ellipse are averaged according to the method given in \cite{Battersby_2025b} to calculate the masses. This ellipse is approximated to be a prolate spheroid in order for the volume density to be calculated from the mass, and the median temperature is taken for the calculation of the sound speed at that size scale ($\approx 3-4$ pc). All further calculations pertaining to the thermal and turbulent Jeans lengths and masses still use Equations 2 and 3 from Section \ref{sec:tjlm}. The CMZ is understood to have steep linewidth-size relations (see: \citealt{Kauffmann_2017b, Shetty_2012}), and the power-law $\sigma \propto R^{0.66}$\footnote{A proportionality constant is not chosen here due to the lack of an appropriate fit for the scale-linewidth relation of the given regions.} (\citealt{Shetty_2012, Kauffmann_2017b}) is used to derive the non-thermal linewidths. See Figure \ref{fig.brickjeans} for comparison between clump-scale and cloud-scale properties.

\section{Mass Segregation Ratio}\label{sec:msts}

We characterize the mass segregation of compact continuum sources identified in the \textit{CMZoom} catalog using the Mass Segregation Ratio (MSR), developed by \cite{Allison_2009a}.

We use this MST-based method to define mass segregation instead of density profiles, mass functions or local density ratios because we choose to focus on source-to-source comparison across the regions. Further, this method does not rely on determining the ``center'' of the cloud, which, for methods such as the mass function, is already assumed to be the location of the most massive stars \citep{Allison_2009a}. This difference in definition complicates comparison with studies using different methods. For specifics on the differences between the methods, please see \cite{Allison_2009a} and \cite{Parker_2015}.

We are unable to calculate the mass segregation ratios for every \textit{CMZoom} survey cloud due to low number statistics (see Appendix \ref{sec:appk} for details). A higher resolution dataset would be needed in order to study these excluded clouds; one of the excluded clouds is G359.484-0.132, which contains only 7 sources in our modified version of the complete catalog.

\subsection{Mass Segregation Ratio: Arithmetic}\label{sec:msrs}

We use the MSR because it is a quantification of the distribution of mass throughout a star-forming cloud that helps to illustrate their morphology. This calculation is complicated by the specific definitions used in this work for both the masses of the compact sources and the derivation of the mass; see Section \ref{sec:err}.  

The arithmetic Mass Segregation Ratio ($\Lambda_{MSR}$), with its standard deviation,

 \begin{equation}
     \Lambda = \frac{\left\langle \lambda_{rand} \right\rangle}{\lambda_{mass}} \pm \frac{\sigma_{rand}}{\lambda_{mass}} \label{msr_eq}
 \end{equation}
is a ratio between the length of the MST for N number of the most massive compact sources in a cloud ($\lambda_{mass}$) and the (arithmetic) average MST length of K sets of N random compact sources ($\left\langle\lambda_{rand}\right\rangle$) \cite{Allison_2009a}, where $\sigma_{rand}$ is the standard deviation of $\left\langle\lambda_{rand}\right\rangle$. The MSR assumes nothing about the specific spatial geometry of a cloud \citep{Parker_2018, Alcock_2019, Dib_2019}. By our definition, an MSR $>$ 1.5 indicates mass segregation, where the more massive compact sources are spatially close together, while an MSR $<$ 0.75 indicates inverse mass segregation, with the more massive compact sources spatially farther apart. An MSR between the bounds shows a uniform mass distribution, with the mass spread evenly throughout the cloud (see Appendix \ref{sec:appk} for details). 

For our sample, we take K = 500 to be sufficient for our calculation of the mass segregation ratio (see Appendix \ref{sec:appk} for an explanation for this choice in K). Some clouds in our sample do not meet the required number of compact sources for us to calculate the MSR with K=500 (see Appendix \ref{sec:appk}) and are not considered in our MSR analysis. For each \textit{CMZoom} cloud with $\geq$ 16 compact sources, we calculate $N_{MST}$ for N = 3 to the last compact source in the cloud, $N_{max}$, and produce an MSR curve.  We plot the MSRs for the \textit{CMZoom} clouds, ranging from N = 3 to N = 20, with K kept constant at 500 in Figure Set 1. Further analysis of these figures is found in Section \ref{sec.firstcase}.
 
According to \cite{Allison_2009a}, the averaging of $\lambda_{rand}$ results in a roughly Gaussian dispersion, which is then quantified by the standard deviation as defined above (Equation \ref{msr_eq}).  We have also followed this method of standard deviation in this work.

The MSR is very sensitive to different assumptions, such as dust temperature, in the compact source-mass estimate. For the catalogs used in this work, and as outlined in \cite{Hatchfield_2020}, the observed Herschel dust temperatures for the clouds range from 20K to 30K. 

Since the MSR is dependent on the construction of the MST, it is sensitive to the inclusion of outlier detections, however, as discussed in Appendix \ref{sec:appk}, our results are minimally impacted. A more complete discussion can be found in Section \ref{sec:err}.

\subsection{Mass Segregation Ratio: Geometric}\label{sec:geom}

We use the Geometric Mass Segregation Ratio  ($\Gamma_{MSR}$) as developed by \cite{Olczak_2011a} and used in \cite{Dib_2019} and \cite{Dib_2018}:

\begin{equation}
    \Gamma = \frac{\langle\gamma_{rand}\rangle}{\gamma_{mass}}
\end{equation}

where $\gamma$ is the geometric mean of the minimum spanning tree lengths, $e_i$:

\begin{equation}
    \gamma = (\Pi_{i=1}^n e_i)^\frac{1}{n}
\end{equation}

and the standard deviation is geometric, as follows:

\begin{equation}
    \Delta\gamma_{rand} = exp\left(\sqrt{\frac{\Sigma_{i=1}^{N_{cs}-1}(\ln{e_{i}}-\ln\gamma_{rand})^2}{N_{core}-1}}\right)
\end{equation}

The K which we use for $\Gamma_{MSR}$ is the same as for $\Lambda_{MSR}$, 500.

\section{Results}\label{sec:results}


\subsection{Arithmetic Mass Segregation in the Clouds}\label{sec.firstcase}

Typically, a region is considered mass segregated if it has an MSR > 1, and inverse mass segregated if it has an MSR < 1. For this work, we choose to enforce stricter criteria to limit the probability that a region is classified as (inverse) mass segregated by chance, where regions with MSR < 0.75 are inverse mass segregated, and regions with MSR > 1.5 are mass segregated. See Appendix \ref{sec:appk} for details of our definition of mass segregation, as compared to the established threshold of MSR $=1$.

\begin{figure*}[h]
\centering
\subfigure{
\includegraphics[width=0.48\textwidth]{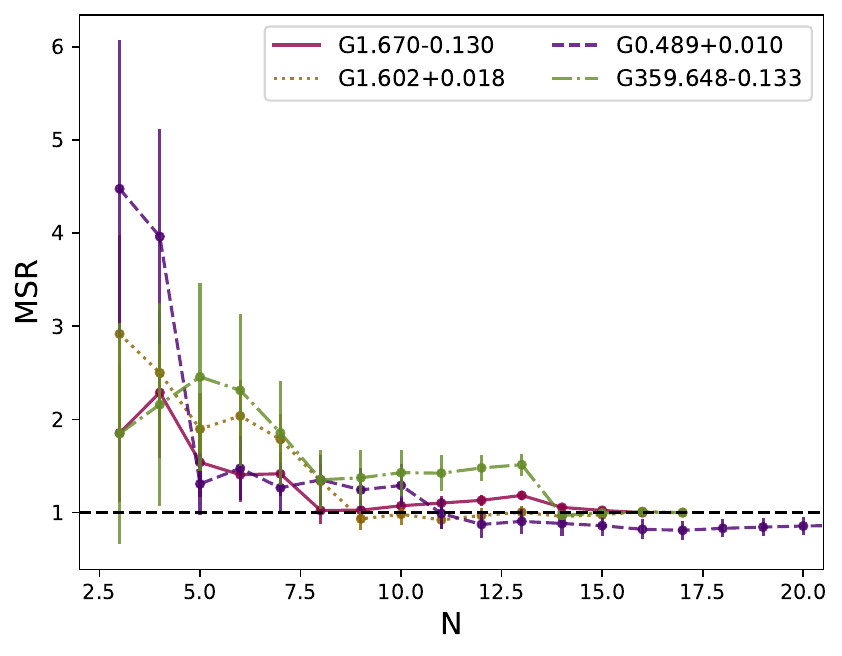}}
\subfigure{
\includegraphics[width=0.48\textwidth]{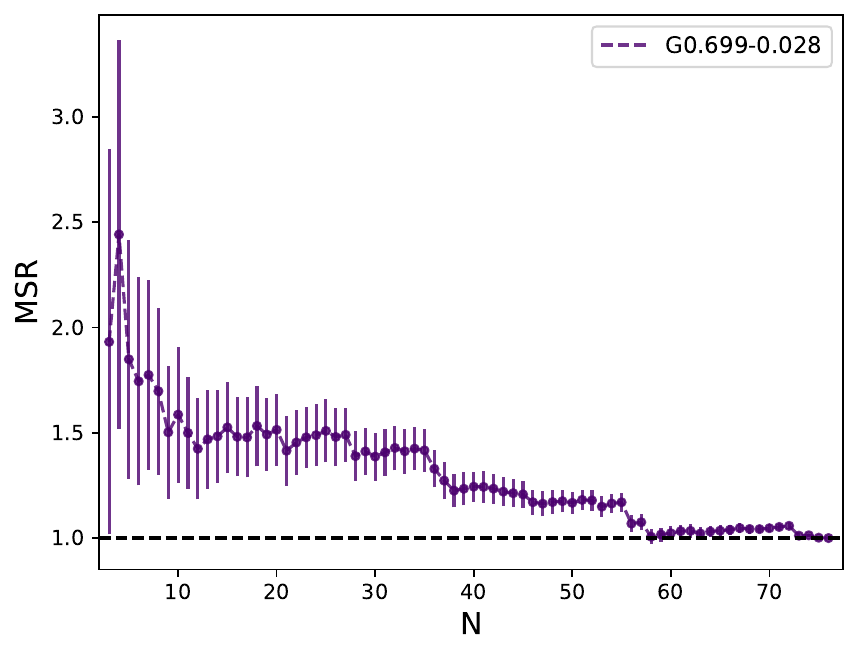}}
\subfigure{
\includegraphics[width=0.48\textwidth]{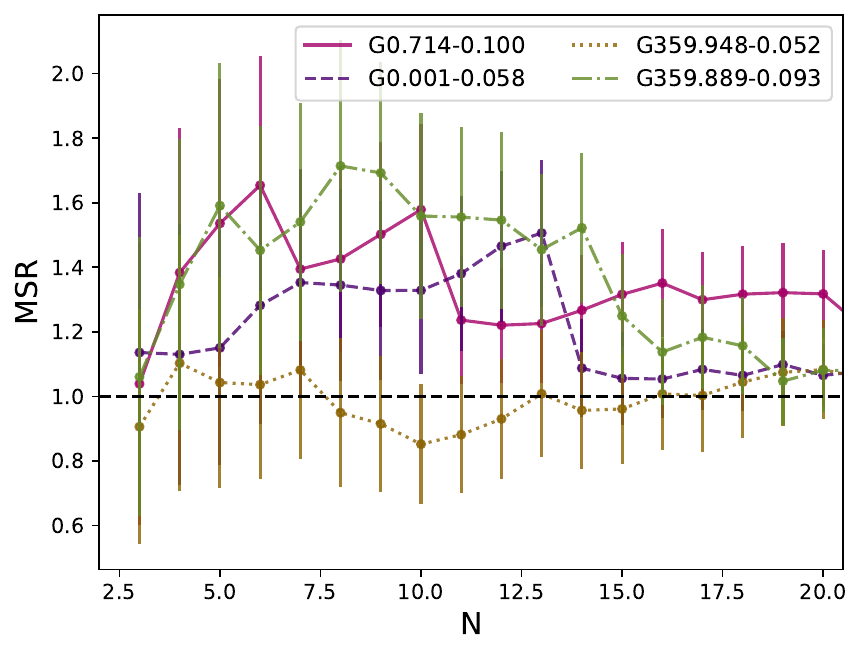}}
\subfigure{
\includegraphics[width=0.48\textwidth]{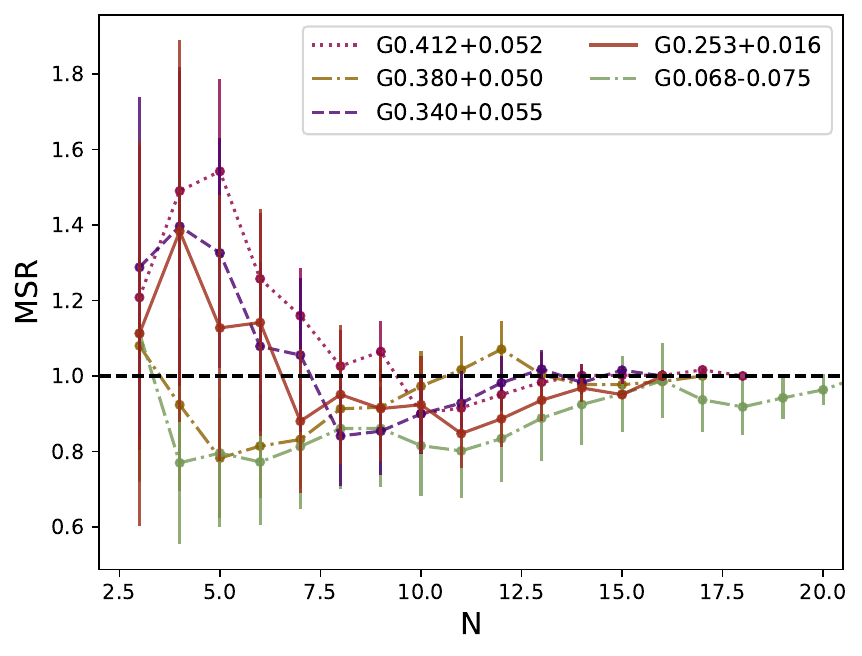}} 
\subfigure{
\includegraphics[width=0.48\textwidth]{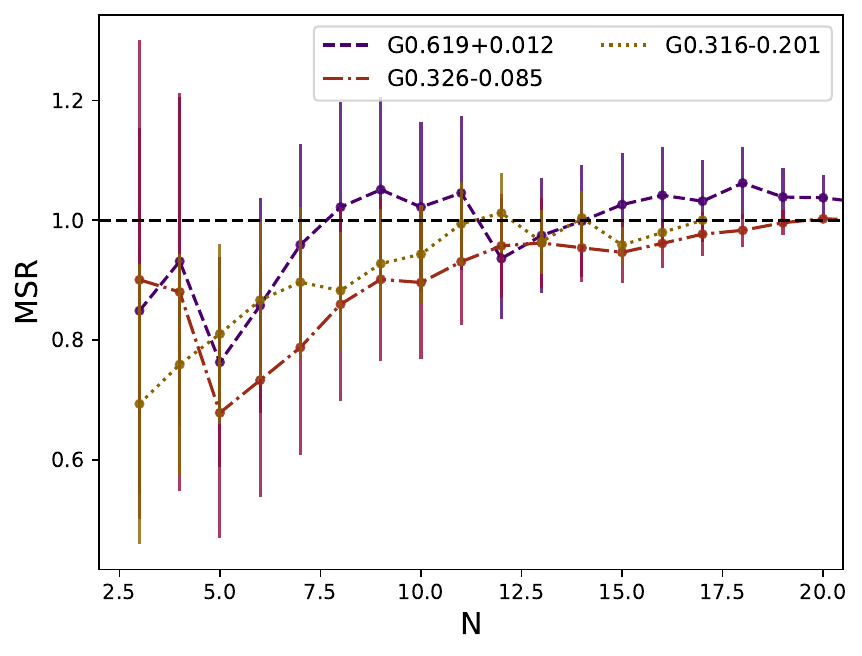}}
\caption{\label{fig.firstcasemsrs} This figure displays the MSR curves for the 17 clouds included in this survey that have at least 16 compact sources detected. (\textit{Top}) The MSR plots for the mass segregated clouds ($\Lambda_3 > 1.5$), with the full plot for G0.699-0.028. N is limited to N = 20 otherwise due to the tendency of the MSR to ``level'' at MSR = 1 for $\frac{N_{total}}{N_{MSR}} \rightarrow 1$. The dotted black line indicates MSR = 1. (\textit{Middle}) The MSR plots for the clouds which show uniform mass segregation ($1.0 < \Lambda_3 < 1.5$), with standard deviation; certain clouds exhibit an increase in MSR at higher N. (\textit{Bottom}) The MSR plots for the inversely mass segregated cloud, G0.316-0.201 ($\Lambda_3 < 0.75$), as well as the clouds which are uniformly distributed at N = 3 with $0.75 < \Lambda_3 < 1.0$.}
\end{figure*}

In the MSR plots (as seen in Figure \ref{fig.firstcasemsrs}), the magnitude of mass segregation is shown on the y-axis and the x-axis shows each consecutive N$_{MSR}$ used in the calculations for the MSR, beginning with N$_{MSR}$ = 3. 

We identify 1 cloud G0.316-0.201 that exhibits inverse mass segregation, or an MSR < 0.75, as shown in the bottom panel of Figure \ref{fig.firstcasemsrs}. 

\label{sec.secondcase}

We consider clouds that exhibit an $\Lambda_3$ between 1.5 and  2.5 (without considering standard deviation) to be mildly mass-segregated (see Appendix \ref{sec:appk} and the top-right and middle panels of Figure \ref{fig.firstcasemsrs}). We identify 3 clouds in this category, G0.699-0.028, G1.670-0.130, and G359.889-0.093. The G359.889-0.093 and G1.670-0.130 clouds show an increase in the degree of mass-segregation with the increase of N, starting from MSR $\approx 1.1$ at N = 3 to MSR $\approx 1.6$ at N = 5, with variation in the curves. The clouds do converge to MSR = 1, as expected. This increase in MSR and subsequent variation does not neatly fit into the definition of mass-segregation, and indicates a complex substructure within the clouds.  The clouds which are shown in Figure \ref{fig.firstcasemsrs} are truncated at N = 20 (unless otherwise stated) so that a crowding effect from the points is avoided.

\label{sec.thirdcase}

The 2 clouds which display slightly stronger mass segregation, G0.489+0.010 and G1.602+0.018, are classified into the third case, see the top left panel of Figure \ref{fig.firstcasemsrs}. This case includes the $\Lambda_3$ values, without considering standard deviation, of $\ge 2.5$. We define the strongest mass segregation to occur when $\Lambda_3$ $\ge 2.5$ within standard deviation. The only region which exhibits an $\Lambda_3$ $\ge 2.5$ within standard deviation is G0.489+0.010 (Dust Ridge E\&F).

\renewcommand\stackalignment{c}
\begin{table*}
\centering
\begin{tabular}{ c c c c c c c c c c c }
\hline
\hline
 Cloud Name & CN & $\Lambda\textsubscript{3}$ & $\Lambda\textsubscript{4}$ & $\Gamma\textsubscript{3}$ & $\Gamma\textsubscript{4}$ & \stackanchor{$\bar{\ell}_{\text{corr}}$}{(pc)} & \stackanchor{$\lambda_{\text{th}}$}{(pc)} & \stackanchor{$\lambda^{\text{H}_2{\text{CO}}}_{\text{turb}}$}{(pc)} & \stackanchor{ M$_{\text{th}}$}{(M$_{\odot}$)} & \stackanchor{M$^{\text{H}_2{\text{CO}}}_{\text{turb}}$}{($10^{+4}$ M$_{\odot}$)}\\ [0.25ex] \hline
 G1.670-0.130 & 16 & 1.85$\pm{0.74}$ & 2.28$\pm{0.71}$ & 1.48$\pm^{3.58}_{0.62}$ & 1.92$\pm^{4.39}_{0.53}$ \\ 

 G1.651-0.050 & 12 &  &  &  &  & 1.72 & 0.30$\pm^{1.01}_{0.23}$ & 6.77$\pm^{2.80}_{1.98}$ & 6.19$\pm^{234}_{6.03}$ & 6.48$\pm^{2.68}_{1.89}$\\

 G1.602+0.018 & 17 & 2.91$\pm{1.06}$ & 2.50$\pm{0.66}$ & 2.93$\pm^{7.95}_{2.93}$ & 2.07$\pm^{9.85}_{5.69}$ & 1.33 &  0.32$\pm^{1.04}_{0.25}$ & 2.12$\pm^{0.88}_{0.62}$ & 6.83$\pm^{257}_{6.65}$ & 2.92$\pm^{1.21}_{0.85}$ \\
 
 G1.085-0.027 & 14 &  & & & & 1.34 &  0.54$\pm^{1.76}_{0.41}$ & 8.13$\pm^{11.5}_{5.75} $ & 12.9$\pm^{488}_{12.6}$ & 11.2$\pm^{15.8}_{7.92}$\\ 
 
 G0.891-0.048 & 2 & & & & & 13.1  & 0.50$\pm^{1.63}_{0.38}$ & 21.4$\pm^{8.86}_{6.27}$ & 12.7$\pm^{479}_{12.4}$ & 99.1$\pm^{41.0}_{29.0}$\\

 G0.891-0.048 & & & & & & & & 13.6$\pm^{5.64}_{3.99}$ &  & 25.5$\pm^{10.5}_{7.48}$\\
 
 G0.714-0.100 & 90 & 1.03$\pm{0.43}$ & 1.38$\pm{0.44}$ & 1.16$\pm^{2.51}_{0.18}$ & 1.44$\pm^{3.50}_{0.61}$ &  1.11 & 0.38$\pm^{1.24}_{0.29}$ & 20.2$\pm^{8.37}_{5.92}$ & 10.2$\pm^{383}_{9.89}$ & 151$\pm^{62.7}_{44.3}$\\

 G0.714-0.100 & &  & & & & & & 10.1$\pm^{4.18}_{2.96}$ &  & 18.9$\pm^{7.84}_{5.54}$\\
 
 G0.699-0.028 & 76 & 1.93$\pm{0.91}$ & 2.44$\pm{0.92}$ & 1.61$\pm^{3.93}_{0.70}$ & 1.93$\pm^{4.50}_{0.64}$ & 0.92 & 0.20$\pm^{0.66}_{0.16}$ & 4.62$\pm^{6.54}_{3.27}$ & 5.43$\pm^{204}_{5.29}$ & 6.37$\pm^{9.01}_{4.51}$ \\

 G0.699-0.028 & & & & & &  & & 6.94$\pm^{9.82}_{4.91}$ &  & 21.5$\pm^{30.4}_{15.2}$ \\
 
 G0.619+0.012 & 23 & 0.84$\pm{0.30}$ & 0.93$\pm{0.27}$ & 0.72$\pm^{1.86}_{0.42}$ & 0.76$\pm^{3.18}_{1.66}$ \\

 
 G0.489+0.010 & 32 & 4.47$\pm{1.59}$ & 3.96$\pm{1.15}$ & 3.50$\pm^{8.53}_{1.53}$ & 3.28$\pm^{8.83}_{2.26}$ &  0.94 & 0.30$\pm^{0.96}_{0.22}$ & 11.2$\pm^{15.8}_{7.90}$ & 7.87$\pm^{297}_{7.66}$ & 42.7$\pm^{60.4}_{30.2}$\\
 
 G0.412+0.052 & 18 & 1.20$\pm{0.40}$ & 1.48$\pm{0.31}$ & 0.96$\pm^{2.68}_{0.76}$ & 1.23$\pm^{4.07}_{1.60}$ & 1.05 & 0.32$\pm^{1.05}_{0.25}$ & 11.9$\pm^{16.8}_{8.43}$ & 9.02$\pm^{340}_{8.79}$ & 45.6$\pm^{64.4}_{32.2}$\\
 
 G0.380+0.050 & 18 & 1.07$\pm{0.40}$ & 0.92$\pm{0.23}$ & 1.06$\pm^{1.12}_{1.00}$ & 0.84$\pm^{2.00}_{0.35}$ & 0.74 & 0.26$\pm^{0.87}_{0.20}$ & 8.82$\pm^{12.5}_{6.24}$ & 7.42$\pm^{280}_{7.23}$ & 27.3$\pm^{38.6}_{19.3}$\\

 G0.380+0.050 & & &  & & & & & 2.94$\pm^{4.16}_{2.08}$ &  & 1.01$\pm^{1.43}_{0.72}$\\
  
 G0.340+0.055 & 16 & 1.28$\pm{0.44}$ & 1.39$\pm{0.42}$ & 1.35$\pm^{2.81}_{0.10}$ & 1.49$\pm^{4.22}_{1.24}$ & 1.07 & 0.43$\pm^{1.39}_{0.33}$ & 15.0$\pm^{21.3}_{10.6}$ & 13.1$\pm^{494}_{12.7}$ & 57.6$\pm^{81.5}_{40.7}$\\
 
 G0.326-0.085 & 21 & 0.90$\pm{0.39}$ & 0.88$\pm{0.33}$ & 0.60$\pm^{3.87}_{2.67}$ & 0.57$\pm^{2.64}_{1.50}$ & 1.12 & 0.72$\pm^{3.08}_{0.16}$ & 26.9$\pm^{38.1}_{19.1}$ & 24.1$\pm^{935}_{0.62}$ & 125$\pm^{177}_{88.2}$\\
 
 G0.316-0.201 & 17 & 0.69$\pm{0.23}$ & 0.75$\pm{0.18}$ & 0.58$\pm^{2.04}_{0.87}$ & 0.75$\pm^{1.81}_{0.30}$ & 0.76 & 0.63$\pm^{2.06}_{0.48}$ & & 18.5$\pm^{700}_{18.0}$ & \\
 
 G0.253+0.016 & 16 & 1.11$\pm{0.51}$ & 1.38$\pm{0.51}$ & 0.90$\pm^{1.46}_{0.55}$ & 1.25$\pm^{1.98}_{0.79}$ & 1.02 & 0.69$\pm^{2.25}_{0.53}$ & 18.6$\pm^{26.4}_{13.2}$ & 22.9$\pm^{866}_{22.3}$ & 45.7$\pm^{64.6}_{32.2}$\\

 G0.253+0.016 & & &  &  & & & & 34.9$\pm^{49.4}_{24.7}$ &  & 301$\pm^{425}_{213}$\\


 
 G0.145-0.086 & 9 & & & & & 1.37 & 0.31$\pm^{1.03}_{0.24}$ & 11.3$\pm^{4.23}_{2.99}$ & 9.21$\pm^{348}_{8.98}$ & 31.6$\pm^{13.1}_{9.27}$\\
 
 G0.106-0.082 & 15 & &  &  &  & 0.58 & 0.31$\pm^{1.02}_{0.24}$ & 10.2$\pm^{4.70}_{3.32}$ & 9.21$\pm^{348}_{8.98}$ & 43.4$\pm^{17.9}_{12.7}$ \\
 G0.070-0.035 & 9 & & &  &  & 2.63 & 0.57$\pm^{1.87}_{0.44}$ & 21.4$\pm^{30.3}_{15.1}$ & 19.1$\pm^{723}_{18.6}$ & 99.1$\pm^{140}_{700}$ \\
 
 G0.068-0.075 & 22 & 1.11$\pm{0.39}$ & 0.77$\pm{0.22}$ & 1.25$\pm^{6.19}_{0.25}$ & 0.80$\pm^{1.82}_{0.36}$ & 0.80 & 0.25$\pm^{0.81}_{0.19}$ & 9.12$\pm^{3.78}_{2.67}$ & 6.91$\pm^{261}_{6.73}$ & 34.9$\pm^{14.4}_{10.2}$\\
 
 G0.001-0.058 & 54 & 1.13$\pm{0.49}$ & 1.13$\pm{0.40}$ & 0.92$\pm^{1.04}_{0.82}$ & 0.88$\pm^{1.26}_{0.62}$ & 0.83 & 0.30$\pm^{0.96}_{0.23}$ & 13.3$\pm^{5.49}_{3.89}$ & 9.44$\pm^{356}_{9.20}$ & 85.7$\pm^{35.5}_{25.1}$\\
 
 G359.948-0.052 & 69 & 0.90$\pm{0.36}$ & 1.10$\pm{0.39}$ & 0.76$\pm^{1.55}_{0.01}$ & 0.98$\pm^{3.35}_{1.37}$ \\
 
 G359.889-0.093 & 48 & 1.06$\pm{0.43}$ & 1.35$\pm{0.45}$ & 1.10$\pm^{1.62}_{0.75}$ & 1.19$\pm^{2.60}_{0.54}$ & 1.01 & 0.33$\pm^{1.06}_{0.25}$ & 14.5$\pm^{5.99}_{4.24}$ & 9.13$\pm^{345}_{8.89}$ & 79.7$\pm^{33.0}_{23.3}$\\
 
 G359.648-0.133 & 17 & 1.84$\pm{1.18}$ & 2.15$\pm{1.08}$ & 1.14$\pm^{3.68}_{1.38}$ & 1.38$\pm^{3.35}_{1.67}$ & 1.43 & 0.83$\pm^{2.72}_{0.64}$ & 13.6$\pm^{5.61}_{3.97}$ & 26.7$\pm^{1008}_{26.0}$ & 12.9$\pm^{5.37}_{3.79}$\\
 G359.615-0.243 & 7 & & & & & 0.90 &  0.59$\pm^{1.92}_{0.45}$ & 4.15$\pm^{1.72}_{1.21}$ & 18.0$\pm^{680}_{17.5}$ & 0.63$\pm^{0.26}_{0.18}$\\
 G359.484-0.132 & 6 & & & & & 2.71 & 0.78$\pm^{2.56}_{0.60}$ & 23.1$\pm^{9.56}_{6.76}$ & 25.0$\pm^{947}_{24.4}$ & 56.5$\pm^{23.4}_{16.5}$\\
\end{tabular}
\caption{
This table shows the arithmetic ($\Lambda_{MSR}$) and geometric ($\Gamma_{MSR}$) MSRs with corresponding uncertainties calculated at $N_{MST}$= 3 and 4 for each eligible \textit{CMZoom} cloud (with source number (CN) $\ge 16$), as well as the calculated $\bar{\ell}_{corr}$ and cloud-scale Jeans lengths and masses for each cloud included in both \cite{Battersby_2025b} and \cite{Walker_2024}; the clouds which are not included are G1.670-0.130, G0.619+0.012, and G359.948-0.052. Certain clouds have multiple row entries in this table since they have multiple linewidths reported in \cite{Walker_2024} $\gamma_{MSR}$ is not explicitly reported, but instead the 1$\sigma$ deviation of $\Gamma_{MSR}$ and $\gamma_{MSR}$. MSRs are calculated using compact sources from our modified version of the complete \textit{CMZoom} catalog. All MSRs have been rounded to 2 decimal places, as well as the Jeans lengths and masses. The turbulent Jeans masses are reported in units of 10$^{+4}$ M$_{\odot}$. } 
\label{tab:mass}
\end{table*}

\subsubsection{Arithmetic Mass Segregation: Robust Catalog} \label{sec:msrrob}

For completeness, we also calculate the arithmetic MSRs for clouds containing at least 12 compact sources in the robust \textit{CMZoom} catalog. We perform this calculation to better characterize the effect of removing low signal-to-noise source detections in our MSR calculations.
As explained in Appendix \ref{sec:appk}, we can not use N = 3 and K = 500 for clouds with only 12 compact sources, and we instead use N = 4 and K = 450. While there are variations in the MSR curves due to fewer compact sources within the clouds, the overall classifications of the analyzed clouds remain unaltered. These values are not reported.

\subsection{Geometric Mass Segregation in the Clouds}
 
The geometric MSR is designed to reduce the overall influence of the outlier compact sources while increasing sensitivity to milder levels of mass segregation, as compared to the arithmetic MSR \citep{Olczak_2011a}. 
The MSR curves that differ 
are seen to have outlier compact sources, such as G0.489+0.010 (see Figure \ref{fig.figuresetex1} for its MST). Outlier compact sources are known to influence the specific shape of the arithmetic curve when included in the K random sets, due to increasing the average MST length. See Section \ref{sec:err} for more details.

\begin{figure}
\centering
\includegraphics[width=0.48\textwidth]{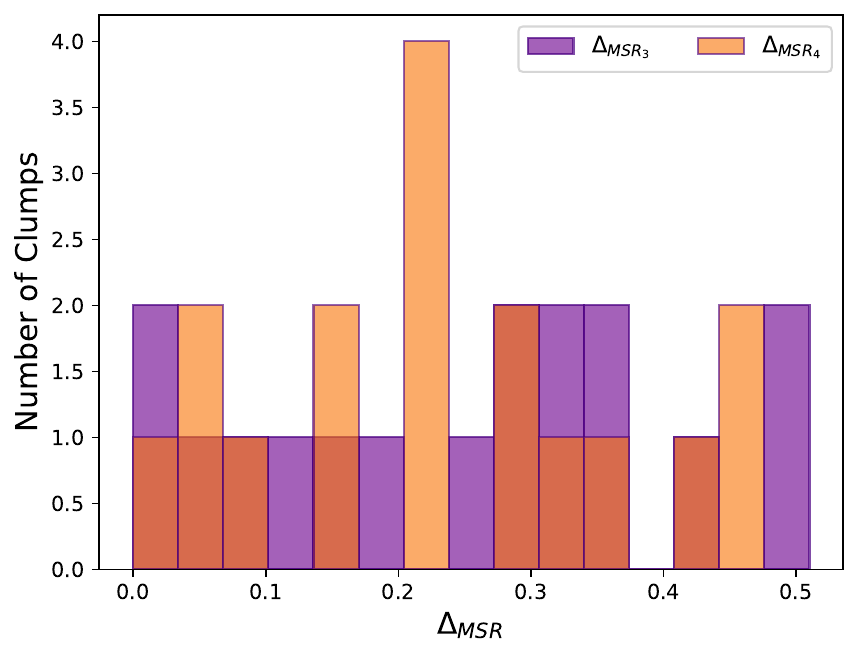}
\caption{\label{fig.diffmsrshist}This histogram shows the difference ($\Delta_{MSR}$) between the arithmetic and geometric MSRs for all 17 clouds that have at least sixteen compact sources, for N = 3 (purple) and N = 4 (orange).}
\end{figure}

Overall, we find that using the geometric MSR calculation does not change the mass segregation classifications for any of the \textit{CMZoom} clouds. 
Shown in Figure \ref{fig.diffmsrshist} is the difference between the geometric and arithmetic methods for N = 3 and N = 4. We find a general agreement between the methods, and outlier compact sources (see Sections \ref{sec:msrs} and \ref{sec:err}) are not so spatially separated as to widely affect the calculated MSR values.

We report the $\Lambda_{MSR}$ and $\Gamma_{MSR}$ for N = 3 and N = 4 for each cloud, along with the standard deviations, in Table \ref{tab:mass}. 

\subsection{Results of the Jeans Analysis}\label{sec:jeans_res}

\begin{figure*}[!h]
\centering
\subfigure{
\includegraphics[width=0.48\textwidth]{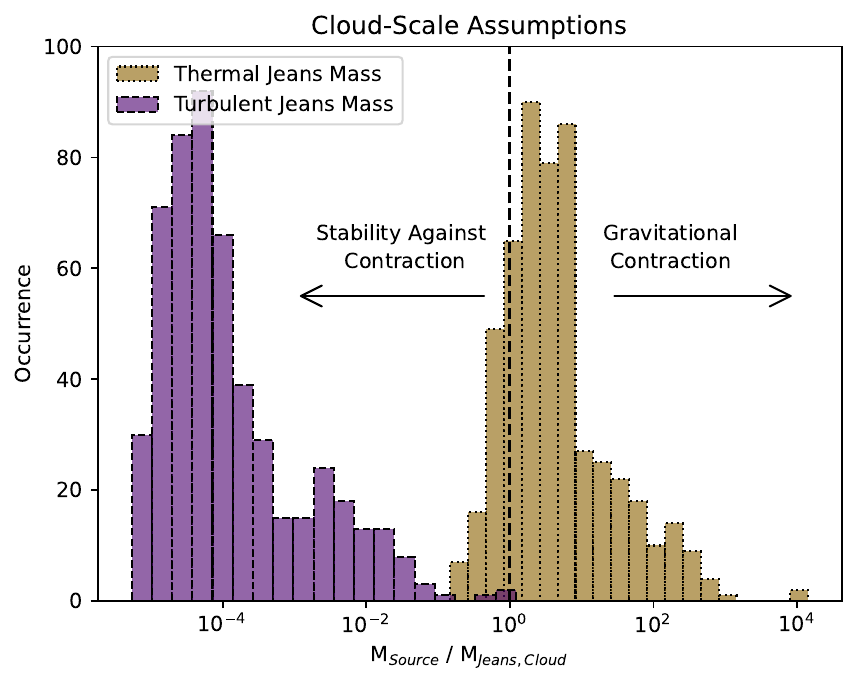}}
\subfigure{
\includegraphics[width=0.48\textwidth]{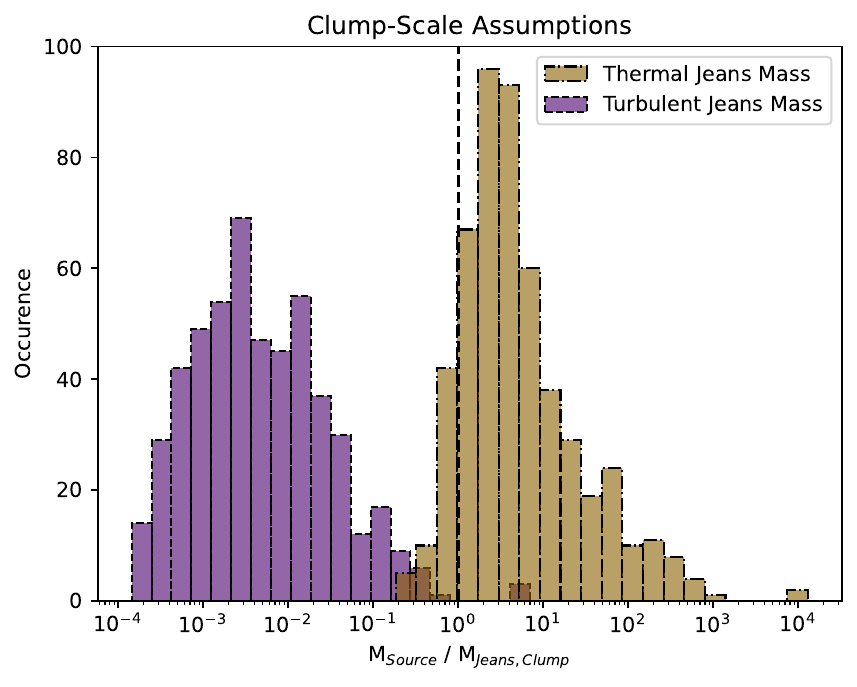}}
\subfigure{
\includegraphics[width=0.48\textwidth]{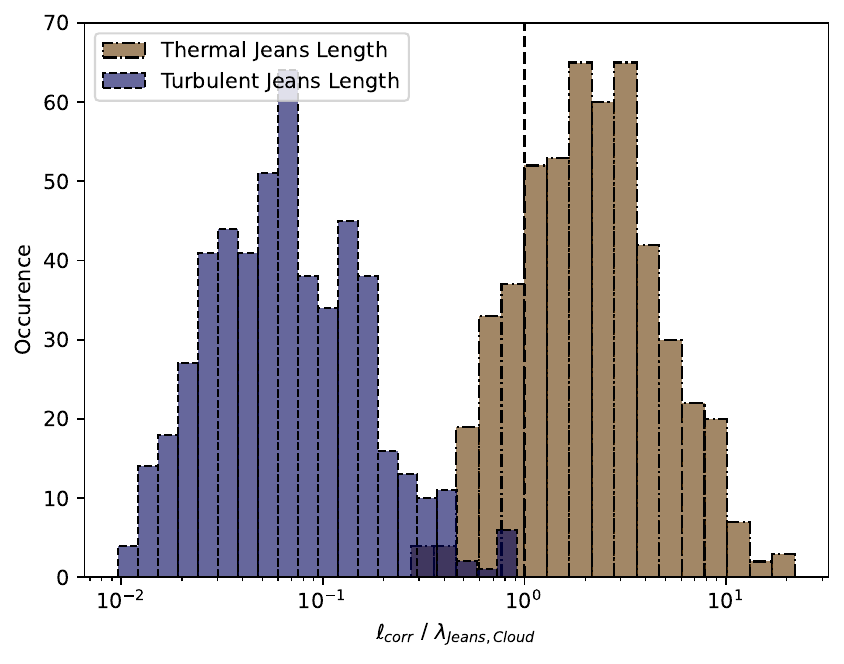}}
\subfigure{
\includegraphics[width=0.48\textwidth]{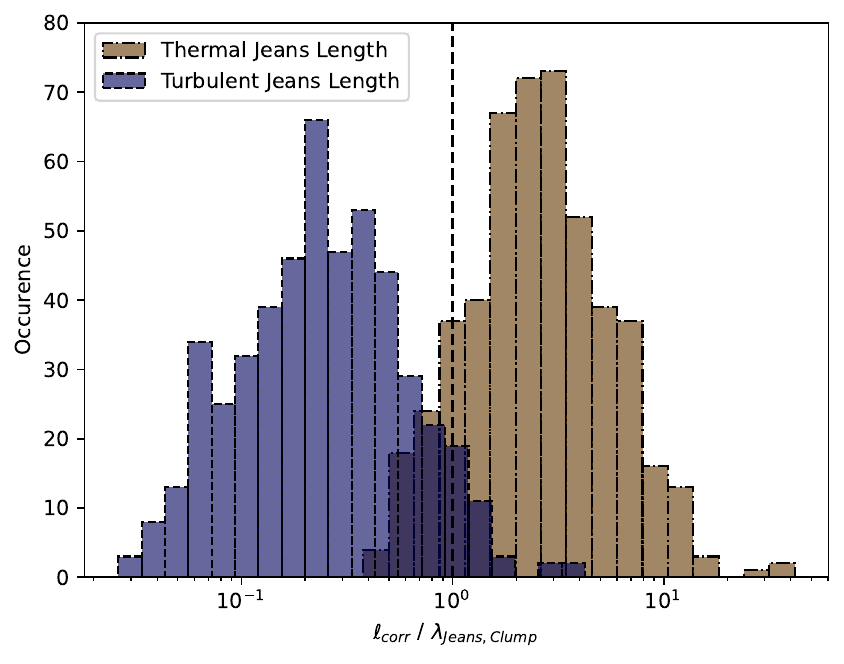}}
\caption{\label{fig.brickjeans} This figure compares compact source separations and masses to the cloud-scale Jeans calculations (lengths and masses) and clump-scale Jeans calculations (lengths and masses). We conclude that the thermal Jeans calculations more closely match the measured values, and also show the potential for gravitational contraction and fragmentation of the sources at small scales while turbulence dominates at large scales. The data is presented as a ratio of values, such that a ratio of 1:1 represents the equilibrium between the gravitational force and the thermal or turbulent pressures of the gas. The x-axes are scaled by log$_{10}$ for concision. (\textit{Top Left}) The histogram of the compact source masses from each cloud as a ratio with the thermal and turbulent Jeans masses for the cloud to which the respective compact sources belong. The dashed line marks equilibrium, a ratio of 1:1. (\textit{Bottom Left}) The compact source separations (deprojected minimum spanning tree edge lengths, $\ell_{\text{corr}}$) in a ratio with the thermal and turbulent Jeans lengths for the cloud to which the respective compact sources belong. (\textit{Top Right}) The  compact source masses in a ratio with  the thermal and turbulent Jeans mass for the respective clump. (\textit{Bottom Right}) The compact source separations in a ratio with the thermal and turbulent Jeans length for the respective clump.}
\end{figure*}

We follow the analysis of \cite{Sanhueza_2019} by comparing the corrected core separations, $\ell_{\text{corr}}$ (not $\bar{\ell}_{\text{corr}}$), of each region with $\lambda_{\text{th}}$ to consider fragmentation at the cloud- and clump-scale. This comparison is performed with ratios of the core properties ($\ell_{\text{corr}}$ and $\text{M}_{\text{source}}$) to the calculated Jeans properties for the clumps and clouds. When the ratio is $\approx 1$, the core properties are similar to the predicted thermal or turbulent Jeans properties, indicating that the cores are consistent with either thermal or turbulent fragmentation. Ratios are performed for both the turbulent Jeans properties (length and mass) and the thermal Jeans properties (length and mass). These ratios can be found in Figure \ref{fig.brickjeans}. 

If we assume a dust temperature of 20 K, we find that the cloud-scale thermal ratios are $\geq 1$ and the turbulent ratios are $\ll 1$ \footnote{The median of the ratio $\frac{\bar{\ell}_{\text{corr}}}{\lambda_{\text{turb}}} \approx 1.07$, and of $\frac{M_{\text{turb}}}{\bar{\text{M}}_{\text{source}}} \approx 2e4$.}.This indicates that the source separations and masses are much smaller than what would be predicted under turbulent fragmentation. At the clump scale with the assumption of the $\sigma \propto R^{0.66}$ power law and the elliptical fit, we find that the turbulent ratios are still largely $\ll 1$, and that using the clump-scale properties in our analysis does not change our results. These calculations include the multiple H$_{2}$CO emission lines detected by \cite{Walker_2024} in 6 of the clouds.

In summary, we find that, for most of the clouds analyzed, $\ell_{\text{corr}}$ is longer than the $\lambda_{\text{th}}$ for the cloud, and not as long as the $\lambda_{\text{turb}}$ for the cloud. Likewise, we find that $M_{\text{th}}$ tends to be smaller than the compact source masses ($\text{M}_{\text{source}}$), and $M_{\text{turb}}$ is generally larger than $\text{M}_{\text{source}}$. At the cloud scale, we observe that the source-separations ($\ell_{\text{corr}}$) are more consistent with thermal fragmentation than turbulent fragmentation, when in the presence of high turbulence present in the CMZ.

These reported results do not include the clump-scale analysis (see Figure \ref{fig.brickjeans} for the smaller-scale results). However, as seen in Figure \ref{fig.brickjeans}, both the cloud-scale and the clump-scale thermal analyses tend to favor fragmentation and gravitational contraction, while the turbulent analyses tend not to.

\section{Discussion}\label{sec:discussion}


\begin{table*}
\centering
\begin{tabular}{c c c c c c c}

\hline
\hline
 Cloud Name & CN & MS & MS$_{\text{BGS}}$ & SF$_{\text{Rob}}$ & SF$_{\text{Amb}}$ & SF$_{\text{3DCMZII}}$ \\ [0.25ex] \hline
 
 G1.670-0.130 & 16 & Yes & Yes & Yes & Yes & --- \\ 

 G1.602+0.018 & 17 & Yes & Yes & Yes & Yes & Yes \\

 G0.699-0.028 & 76 & Yes & Yes & --- & --- & Yes$^{*\diamond}$  \\

 G0.489+0.010 & 33 & Yes$^{\dagger}$ & Yes$^{\dagger}$ & Yes$^{*}$ & Yes$^{*}$ & Yes$^{*}$ \\

 G359.648-0.133 & 17 & Yes & Yes & No & No & --- \\

 G0.380+0.050 & 18 & Un. & Yes & Yes$^{*}$ & Yes & Yes$^{*\ddagger}$ \\

 G359.889-0.093 & 48 & Un. & Yes & Yes$^{*}$ & Yes$^{*}$ & Yes$^{\ddagger}$ \\

 G0.326-0.085 & 21 & Un. & Yes & No & Yes & --- \\
 
 G0.619+0.012 & 23 & Un. & Un. & Yes$^{*}$ & Yes$^{*}$ & --- \\

 G0.001-0.058 & 55 & Un. & Un. & Yes & Yes & Yes \\
 
 G0.412+0.052 & 18 & Un. & Un. & No & Yes & Yes \\

 G0.714-0.100 & 90 & Un. & Un. & No & Yes$^{*}$ & Yes$^{*}$ \\

 G0.340+0.055 & 16 & Un. & Un. & No & No & Yes$^{\diamond}$\\

 G359.948-0.052 & 69 & Un. & Un. & --- & --- & --- \\

 G0.068-0.075 & 23 & Un. & No & Yes & Yes & Yes \\

 G0.253+0.016 & 17 & Un. & No & No & Yes & Yes$^{\diamond}$ \\
 
 G0.316-0.201 & 17 & No & Un. & No & No & --- \\

 \hline
\end{tabular}
\caption{\label{tab:sfr} In this table, we compare mass segregation for the 17 applicable \textit{CMZoom} regions (with CN $\ge 16$) with the presence of star-formation tracers as derived and shown in Tables 2 and 3 of \cite{Hatchfield_2024} (the columns labeled ``SF$_{\text{Rob}}$'' and ``SF$_{\text{Amb}}$'' here) as well as in Table 4 of \cite{Battersby_2025b} (``SF$_{\text{3DCMZII}}$''), with ``SF'' indicating that the region is known to be star-forming. $\dagger$ indicates clouds with MSR $>$ 4. $*$ indicates star-formation rates (SFR) $\ge$ 1.0e-02 $M_{\odot} yr^{-1}$. $\ddagger$ indicates clouds where the SFR is calculated using a uniform T of 50 K rather than the Herschel Hi-GAL temperature ($\approx 20$ K). $\diamond$ indicates SFR estimates calculated using free-fall time. Distinct MSRs $>$ 1.5 are represented by ``Yes'', MSRs between 0.75 and 1.5 with ``Un.'' (``Uniform''), and MSRs $<$ 0.75 with ``No''. MS$_{\text{BGS}}$ is the MSR as derived from the compact source masses with the background flux subtracted, see Section \ref{sec:err}. G0.699-0.028 and G359.948-0.052 are excluded from \citep{Hatchfield_2024} due to uncertainties in the mass estimates, and G359.948-0.052 is excluded from \cite{Battersby_2025b} due to a lack of background-subtracted structure.}
\end{table*}

As shown in Figure \ref{fig.firstcasemsrs}, the majority of the regions analyzed in this paper exhibit either inverse mass segregation or a nearly uniform mass distribution, with only 5 classified as mass segregated. This indicates that mass segregation is not necessarily common in these regions of the CMZ at this resolution and size scale.

\subsection{Mass Segregation and Star Formation} 

As discussed briefly in the Section \ref{sec:intro}, the CMZ might be underproducing stars given the amount of dense gas it has overall. \cite{Battersby_2025b} finds that the calculated SFRs, while low for the entire CMZ, are expected for the gas contained in the clouds which are currently forming stars, and concludes that much of the gas and dust in the CMZ is simply not engaged in star-formation. We use the SFRs calculated in \cite{Hatchfield_2024} to determine the relationship between the mass segregation of our clouds and their star formation activity, as \cite{Hatchfield_2024} covers a complete sample of clouds throughout the CMZ. 

We find that four clouds consistently show agreement between observed mass segregation and detected star-formation with the non-background subtracted compact sources, and that one cloud shows inverse mass segregation and no star-formation tracers. With the background subtracted compact sources, seven clouds show both mass segregation and star formation, while two clouds show neither.

There is no obvious correlation between the mass segregation observed in this work and the evolutionary stage of the clouds analyzed. G0.489+0.010 exhibits the highest degree of mass segregation in this work, while G0.380+0.050 exhibits a roughly uniform level of mass segregation, and G0.412+0.052 (Dust Ridge D) a slightly higher MSR of $\approx 1.20$; the first two clouds are considered to be evolutionarily young due to the presence of star formation tracers \citep{Lu_2019a} \citep{Barnes_2019, Walker_2017} and the lattermost even younger \citep{Barnes_2019}. G0.699-0.028, which presents a lower and more complex MSR than G0.489+0.010, is likely more evolved due to a large population of protostellar cores \citep{Ginsburg_2018a}. The mostly quiescent G0.253+0.016 cloud, which displays uniform mass segregation, is considered to be at an early evolutionary stage \citep{Walker_2021, Mills_2015, Immer_2012}, with a steeper CMF than that of the actively star-forming G0.699-0.028 \citep{Kinman_2024}. G1.602+0.018 displays mass segregation but has not been extensively studied as to its current evolutionary state. The other clouds in this work are too ambiguous with respect to their mass segregation values, and provide little to no statistical significance to a correlation between mass segregation and star formation (or mass segregation and evolutionary stage).

In summary, we do not see clear evidence of a correlation between mass segregation and SFR in this work. We find that mass segregation is not the best indicator for current star-formation within the clouds, and that clouds which are actively star-forming are not necessarily more likely to be primordially segregated.

\subsection{Possible Sources of Error}\label{sec:err}

There are multiple factors that can contribute to uncertainty in our analysis that come from both the data and our methods.  


\begin{enumerate}

\item \textbf{Sources in 2D} One potential source of error is our definition of a ``compact source''. These are defined based on their relative column density compared with the background continuum. However, these clouds are observed as two-dimensional on the sky, and are three-dimensional in reality. This can allow multiple compact sources to ``stack'' on top of each other in the line-of-view. Further, the masses of each compact source are derived from the overall flux. Multiple compact sources that are stacked on top of each other will appear to have a higher flux than if they had been spread out individually. This has the potential to affect not only the counted number of compact sources in each cloud but also the masses of those compact sources. This might explain the unusually high masses of the compact sources in the G0.699-0.028 cloud. Further, if the compact sources are possibly under-counted due to the method of observation used, the MSTs drawn from the observed compact sources do not properly represent the morphology of the cloud, and this in turn affects the derived MSRs. Differences in dust temperature also have the potential to affect the calculated MSRs (as briefly discussed in Section \ref{sec:msrs} and reported in \citealt{Hatchfield_2020}). 

\item \textbf{Imaging Artifacts} Furthermore, the noise present in the SMA data can result in possible false positive detections, particularly along the defined edges of the regions. Because of this, as previously stated in Section \ref{sec:data}, we use a version of the complete catalog with the edge sources removed. We also acknowledge that sources might exist beyond the boundaries of our defined regions, however, the \textit{CMZoom} observations were designed to cover the densest clouds in the CMZ where star formation would be located, so we consider it unlikely that any significant sources are outside of the regions investigated here.

The G0.699-0.028 and G359.948-0.052 regions suffer from imaging artifacts and can be expected to have higher local rms noise levels as a result. Since the \textit{CMZoom} catalog accounts for the local noise at the location of each leaf, we would expect these regions to be pruned more stringently for inclusion in either the robust or complete catalogs than in other regions that may lack these imaging artifacts.

\item \textbf{Outlier Sources} We find that very few regions have statistical outliers, where there are MST  edge lengths that vary significantly from the mean edge length, while some regions contain none. Of the sources which may be outliers, as often as not, they are centrally located within the region, or otherwise located such that their removal from the set results in the creation of more severe outliers. For our analysis, we are mainly concerned with potential false detections from noise, which may produce ``massive'' outliers that are not centrally located. This is the case with regions G1.602+0.018 and G359.648-0.133 in the complete catalog, where the massive outliers lie at the edge of the map. The subsequent removal of these edge sources in the reduced version of the complete catalog allows these regions to become more mass segregated.

\item \textbf{Complete vs. Robust Catalog} As mentioned in Section \ref{subsec:cat}, we primarily use the complete \textit{CMZoom} catalog in this work (although we do briefly discuss the differences in results with the \textit{CMZoom} robust catalog, in Section \ref{sec:msrrob}). Our choice of catalog can potentially impact our resulting MST analysis. The complete catalog likely contains ``false positive'' detections from the noise in the observations, whereas the robust catalog may be removing real, low-brightness sources. These differences propagate out to the results from our analysis, and the truth of the clouds can be assumed to lie somewhere in-between. \cite{Hatchfield_2020} consider the completeness of both catalogs, and find that the complete catalog is 95\% complete at 50 M$_{\odot}$, while the robust catalog is 95\% complete at 80 M$_{\odot}$. Furthermore, as discussed briefly in Section \ref{sec:data},  there is unlikely to be very many false positives among the lower mass cores of the complete catalog, with the robust catalog already containing most of the mass within the regions.
\cite{Hatchfield_2020} compared the SMA data for a few key regions with corresponding ALMA data in order to better understand the veracity of the catalog sources, with respect to the filamentary structures. The authors find that there is good agreement between the images, while the G0.699-0.028 SMA image contains some sources not found in the ALMA image. This discrepancy could be due to the fact that the G0.699-0.028 region contains both the highest noise and the highest flux found within the \textit{CMZoom} catalog. 
Missing detections or including false detections from the catalog will impact our results, however the \textit{CMZoom} catalog included high flux thresholds in the creation of the catalog and considered the influence of edge sources in order to mitigate the impact of false detectons. Most missing detections will be lower mass sources that would have less impact on the MSR calculation. Additionally, we show that the MSR classifications for each region changes minimally when using the robust catalog, as shown in Section \ref{sec:msrrob}.

\item \textbf{Measurement Errors} When we use the background subtracted mass for each source, the position of the most massive compact source in a cloud may change. This affects the MSR directly. We observe that some clouds, such as G359.948-0.052 near Sgr A*, become more mass segregated with the background subtracted masses. The majority of the clouds analyzed do not exhibit any significant difference in the MSR values when the background subtracted masses are used compared to the MSR values when the background is not subtracted from the mass estimations. The clouds for which there is a significant difference are shown in Figure \ref{fig.bgscomp}. Based on this evaluation, we do not consider the use of the background subtracted masses to chnage our conclusions regarding our MST-based analyses. 

Additional sources of error come from the analysis with the thermal and turbulent Jeans lengths and masses. One source of error is the median dust temperature used in the calculation of the thermal Jeans lengths and masses, and stated in Section \ref{sec:tjl} to be roughly $10 \%$. This error is further complicated by systematic effects including the variations of temperatures along the line of sight, within a beam, or variations between the small scales of the SMA and the larger scales of Herschel (see: \citealt{Battersby_2020, Battersby_2011}). Higher resolution studies of the CMZ, such as \cite{Mills_2013} and \cite{Ginsburg_2016}, find higher temperatures associated with smaller-scale structures located within the clouds.

Another source of error belongs to the densities from the Herschel Hi-GAL table, which are considered to have uncertainties of a factor of 2 (also discussed in Section \ref{sec:tjl}). The last source of error is from the compact source masses themselves, as discussed in \cite{Battersby_2020}, which are considered also to have uncertainties of roughly a factor of 2 (see \citealt{Battersby_2020} for more detail). 

\item \textbf{Reliability of the \textit{CMZoom catalog}} The compact sources are treated as point sources in this work, which raises concerns over the validity of calculations involving their separations, particularly given that no two sources can physically be closer than their radii. The dendrogram leaves have hard boundaries which do not overlap, resulting in a limit to how close the leaves can be to each other. This could potentially result in false negatives, as in a hypothetical where projection causes two otherwise distinct sources to be treated as one. Hypothetical sources which are close enough to each other to raise concerns over projection and may be considered one massive object by the dendrogram algorithm would be likely to contribute toward shorter MST edge lengths. We do not consider this possible error to significantly affect our MSR analysis, since a hypothetical massive source which causes a region to be considered mass segregated, if instead considered to be multiple smaller sources, would nonetheless contain a larger fraction of mass than the other sources in the region, and therefore, the mass is still segregated in that part of the region. 

\end{enumerate}



\section{Conclusions}

The Central Molecular Zone has long been considered to have unusually low levels of current star-formation given its extreme physical conditions. The \textit{CMZoom} survey has allowed us to investigate mass segregation across the densest star forming clouds in the CMZ in a consistent and comparable manner, in order to better understand the differences and similarities between the clouds, rather than focusing on only one cloud at a time.

We use MST-based analyses to quantify the spatial distribution and mass segregation of compact sources identified in the \textit{CMZoom} survey. We also investigated how mass segregation correlates with the presence of star formation tracers and thus the evolutionary stage of a cloud. We summarize our findings below:

\begin{enumerate}
    \item We compare the MSRs of the clouds to each other to better understand the CMZ as a whole, rather than cloud by cloud as it has been considered previously. We find that one-third of the clouds surveyed exhibit mass segregation, while the remaining two-thirds exhibit either inverse mass segregation or uniformity. 
    \item We calculate both the arithmetic and geometric mass segregation ratios of the clouds included in the \textit{CMZoom} survey with more than 15 compact sources. We find that the geometric MSRs generally present the same curves as the arithmetic MSRs, as expected, and in clouds with outliers, the effect on the MSR is small. 
    \item We find that, despite the widespread presence of turbulence in the CMZ, the clouds included in this work appear to be fragmenting according to the thermal Jeans length and mass, at the observed source resolution of $\approx 0.1$ pc. This result is consistent with those found in other CMZ studies that observed fragmentation at $\leq 0.1$ pc scales (\citealt{Lu_2020, Kruijssen_2019, Pokhrel_2018}). 
    \item We briefly consider a possible link between the MSRs calculated in this work and the star-formation rates presented in \cite{Hatchfield_2024}, as well as the presence of star formation tracers as an evolutionary indicator for the clouds. We do not find a strong correlation between the MSRs and the presence of star formation. 
\end{enumerate}

This is the CMZ-wide study of the spatial distribution and mass-segregation of compact sources, where previous studies have considered individual clouds. Ultimately, we find that there is no strong correlation between the mass segregation for star-forming regions in the CMZ and their evolutionary state at 0.1 pc scales; however, higher resolution studies towards a large and representative sample of regions will be needed to see if this is still true at core (~1000 au) scales.


\section{Acknowledgments}

We acknowledge the use of the \textit{CMZoom} Survey, first published in 2020, without which this work could not exist. 
S. Schuler, J. Wallace, and C. Battersby gratefully acknowledge funding from the National Science Foundation under Award No. 2108938. S. Schuler and C. Battersby were also supported by the National Science Foundation Award No. 2206510 and J. Wallace and C. Battersby were supported on Award No. CAREER 2145689. C. Battersby also gratefully acknowledges support from the National Aeronautics and Space Administration through the Astrophysics Data Analysis Program under Award ``3-D MC: Mapping Circumnuclear Molecular Clouds from X-ray to Radio,” Grant No. 80NSSC22K1125.   
S.Z.\ acknowledges support from the Strategic Priority Research Program of the Chinese Academy of Sciences (CAS) Grant No.\ XDB0800303. 
RAG acknowledges funding support from: NASA ADAP awards NNX11AD14G, NNX13AF08G, NNX15AF05G, and NNX17AF24G; NSF AAG grants 1636621, 1812747, and 2107705; NASA-USRA SOFIA grants 05-0181, 07-0225, and 08-0181; NASA-JPL/Caltech Spitzer grants 1373081, 1424329, and 1440160, and Herschel grant 1489384. 

\appendix \label{sec:apps}

\section{Discussion of K Values, and the Arithmetic MSR}\label{sec:appk}

As stated in Section \ref{sec:msrs}, we choose a K value of 500. This choice comes from the behavior of the Arithmetic MSR when calculated on randomly uniformly distributed clouds. \cite{Allison_2009a} state that there is a roughly Gaussian dispersion associated with the mean lengths of the random MSTs (quantifiable by the standard deviation $\left\langle \ell \right\rangle \pm \sigma$). This is due to the specific random sets of compact sources that are chosen and averaged to make the numerator of the ratio. With an increase of K, especially if the sets are randomly chosen without replacement, a greater fraction is taken of the total number of random sets available. Each time that the MSR calculation is run with replacement sampling, it is possible for the same random sets to be sampled again, which can produce a less accurate calculation of the MSR.
 
\begin{figure}[h!]
\centering
\subfigure{
\includegraphics[width=0.48\textwidth]{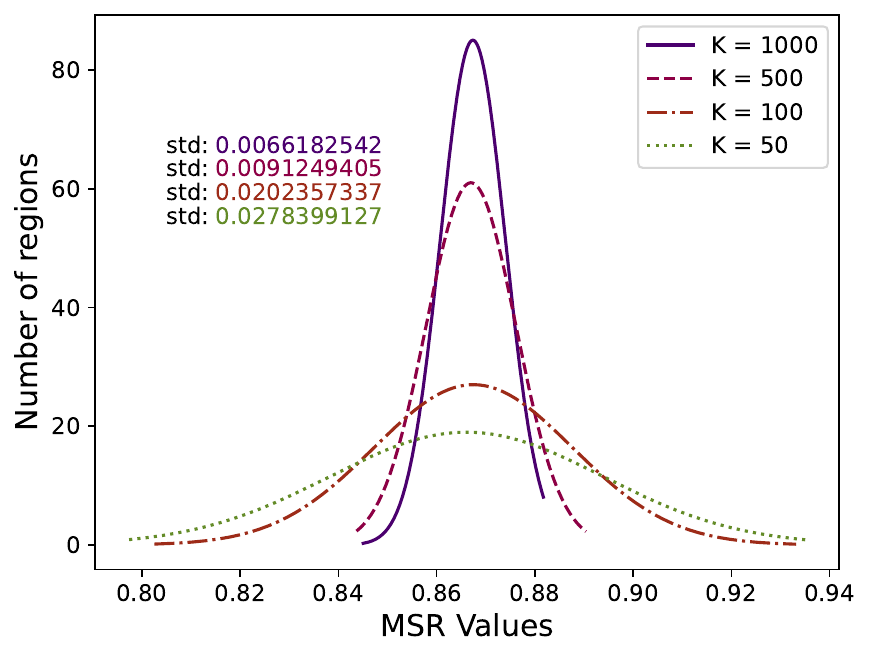}}
\subfigure{ 
\includegraphics[width=0.48\textwidth]{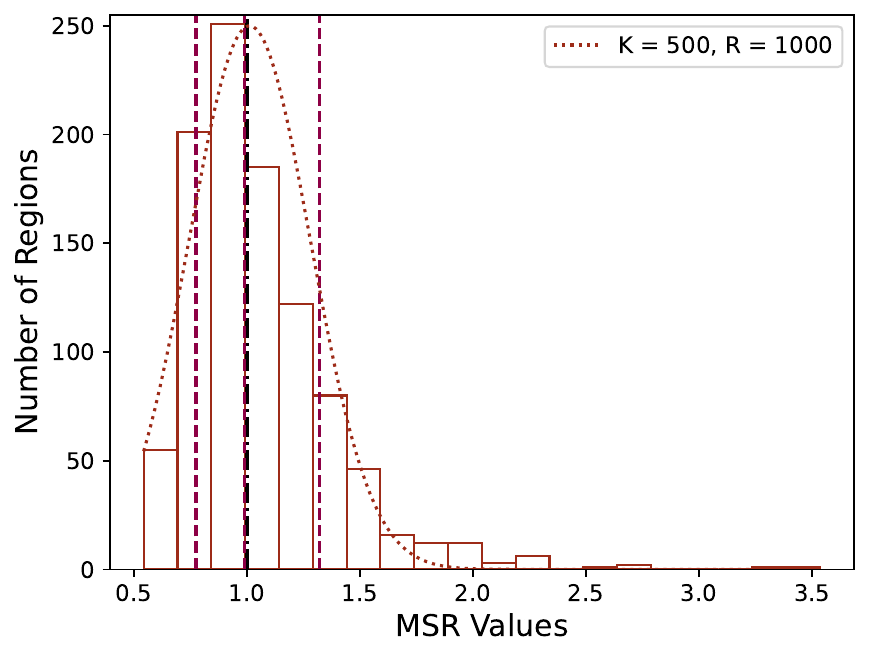}}
\caption{\label{fig:kvalues} (\textit{Left}) The distribution of MSR values for an ensemble of randomly uniformly distributed simulated clouds showing a decrease in the standard deviation with an increase in K, with the standard deviation listed on the left-hand side of the panel. The same cloud was iterated over 200 times for each chosen K with N$_{MST}$ = 4 and a total compact source number of 20. The dashed lines are Gaussian approximations of the spread of the MSR values for each K. The figure is centered at an MSR of 0.87 because that value is where the majority of the calculated MSRs naturally fell --- 0.87 is the median MSR for this particular randomly uniformly distributed cloud. (\textit{Right}) The log-normal nature of the MSR distribution for 1000 randomly uniformly distributed clouds, at K = 500. The dashed vertical lines mark the 1$\sigma$ percentages, and the dashed-dotted black line marks MSR = 1.0.}
\end{figure}

When sampling the random sets without replacement, it is important to only sample up to the number of available unique combinations. Therefore, setting a specific K will dictate the smallest N$_{MST}$ and lowest compact source number (CN) that can be used. We find that a K value of 500 is adequate for our analysis in this work, as it allows us to have a CN as low as 16 and an N$_{MST}$ as small as 3. See the left hand panel of Figure \ref{fig:kvalues} for the comparison between K values for the same randomly generated cloud with 20 compact sources and $N_{MST}$ = 4. 
\par \cite{Allison_2009a} recommend a K of 500 or 1000 for low N$_{MST}$ and a K of at least 50 for high N$_{MST}$. This is due to the fact that as N$_{MST}$ approaches the total CN of a given uniformly distributed cloud, the random MST lengths do not differ from each other greatly, allowing for there to be smaller deviations between them. This can be seen in Table \ref{tab:stdv}, where the standard deviation of a cloud with 25 compact sources, N$_{MST} = 4$, and K $= 500$ is similar to the same cloud with N$_{MST} = 13$ and K $= 50$. The same holds for other clouds of different sizes. We choose K $= 500$ for our clouds regardless of N$_{MST}$ to obtain results of the same or better precision. 

The established definition of inverse mass segregation describes when a cloud exhibits MSR values lower than 1.00, and is observed only to fit HMSF 0. By our definition, these values between 0.75 and 1.00 fall within the $1\sigma$ deviation of a uniformly distributed cloud (see Appendix \ref{sec:appk}), and are neither truly inversely mass segregated nor mass segregated. This can be seen in the right panel of Figure \ref{fig:kvalues}, where the plotted MSR distribution for 1000 unique, randomly uniformly distributed clouds at a K of 500 shows that, for randomly assigned mass values within the simulated regions, the MSR follows a log-normal distribution with the majority of the MSR values falling between 0.75 and 1.50. We therefore take these values as bounds for MSRs indicative of uniformly distributed regions.

\subsection{Confidence of the Arithmetic MSR}\label{sec:amsr}
The mass-segregated clouds, or the MSR values themselves, are fairly confident per region in the \textit{CMZoom Survey}. This is tested by taking a region at a certain N  and calculating the MSR for that region at that N 1000 separate times. These 1000 MSR values are then looked at to see how many fall into a different classification than where the region was originally placed. All regions at N = 3 and N = 4 consistently have MSR values within the same classification. At higher N values, the morphology of the particular region plays a part in what fraction of the 1000 iterations will fall into a different class at that N, however, the MSR naturally moves through classes as it approaches 1. This suggests that any effect on the MSR by possible outliers is negligible.

\cite{Maschberger_2011} suggests the use of median edge lengths rather than mean edge lengths in the calculation of the MSR in order to minimize the influence of outliers. In our tests of \cite{Maschberger_2011}'s method compared to \cite{Allison_2009a}'s method, we find that at N = 3 for the reduced catalog, the “median” MSR, when iterated 1000 times, tends to have greater standard deviation, as can be expected by the use of the median. Additionally, for G359.648-0.133, Maschberger’s method results in 11 out of 1000 iterations (1.1\%) falling into a different category.

We test the sensitivity of the arithmetic MSR to ``leave-one-out'', where each compact source is excluded from the sample. We consider the use of the complete catalog for two regions: G1.602+0.018 and G0.489+0.010, due to the complete catalog's higher population of cores and potential outliers. The former region contains a “massive” compact source close to the edge of the region, which is included in the minimum spanning tree for the most massive sources. When this source is removed, the MSR changes, otherwise, the MSR is consistently uniform with the removal of the other sources. The latter region contains two sources near to its edges, however, these sources are not massive, and the exclusion of each source does not affect the MSR. When these two sources are excluded one at a time, the MSR changes to reflect the average shortening of the random minimum spanning trees; this change is not significant overall due to the high value of the MSR to begin with.

\begin{table*}[h!]
\centering
\begin{tabular}{c c c c}
\hline
\hline
 CN & N$_{MST}$ & K Value & $\sigma$ \\ [0.25ex] 
\hline
25 & 4 & 50 & 0.0278 \\

25 & 4 & 100 & 0.0202 \\

25 & 4 & 500 & 0.0091 \\

25 & 4 & 1000 & 0.0066 \\

25 & 13 & 50 & 0.0099 \\

25 & 13 & 100 & 0.0065 \\

25 & 13 & 500 & 0.0029 \\

25 & 13 & 1000 & 0.0022 \\

200 & 32 & 50 & 0.0080 \\

200 & 32 & 100 & 0.0065 \\
 
200 & 32 & 500 & 0.0026 \\
 
200 & 32 & 1000 & 0.0018 \\
 
200 & 100 & 50 & 0.0037 \\
 
200 & 100 & 100 & 0.0025 \\
 
200 & 100 & 500 & 0.0011 \\
 
200 & 100 & 1000 & 0.0008 \\
\hline 
\end{tabular}
\caption{In this table are reported the standard deviations for unique randomly uniformly distributed simulated clouds of a certain CN and N, with varying Ks. The unique clouds have been sampled 200 separate times in order to determine the standard deviations.}\label{tab:stdv}
\end{table*}

\section{MSRs for Background-Subtracted Sources}\label{bgsmsr}

\begin{figure*}[htb!]
\centering
\subfigure{
\includegraphics[width=0.48\textwidth]{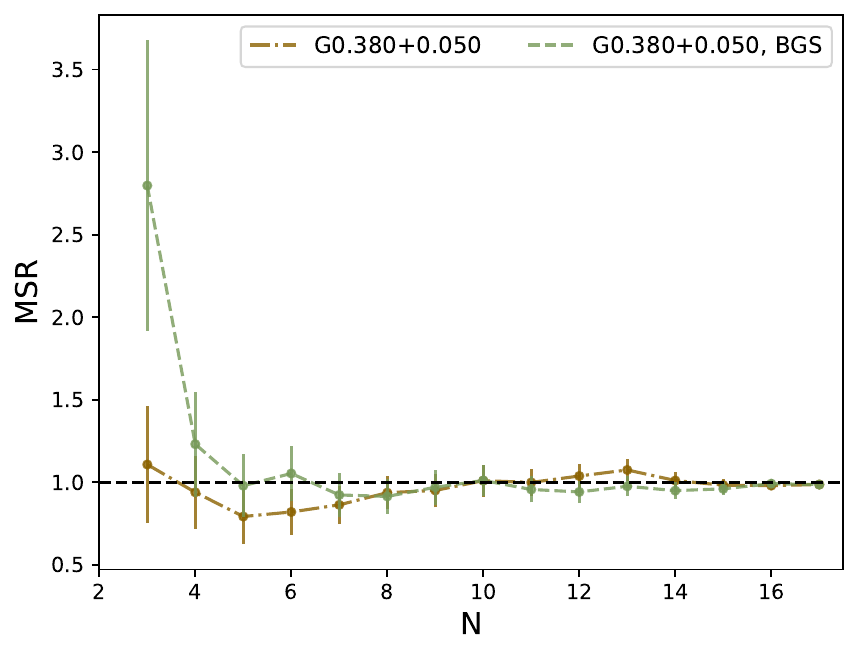}}
\subfigure{
\includegraphics[width=0.48\textwidth]{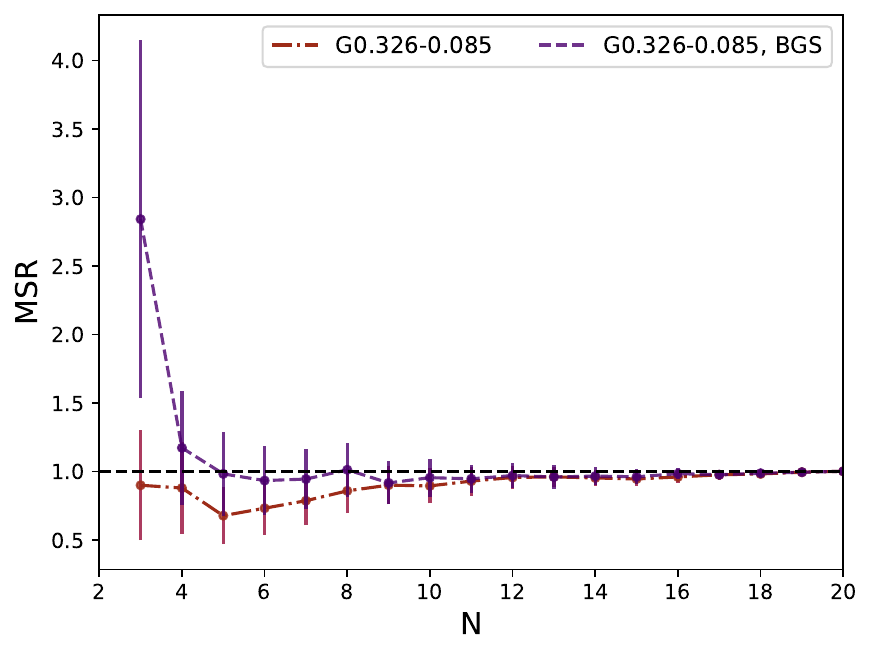}}
\subfigure{
\includegraphics[width=0.48\textwidth]{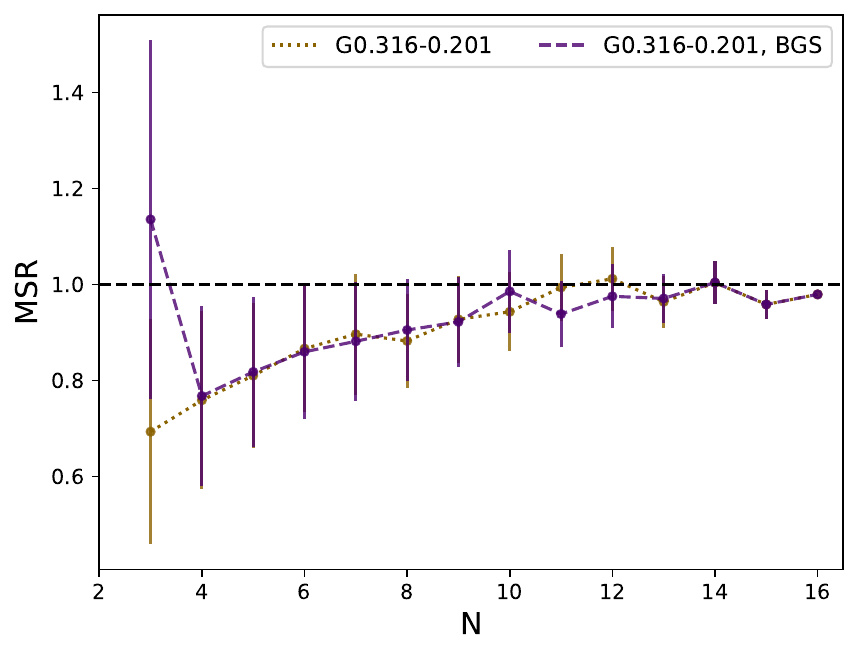}}
\subfigure{
\includegraphics[width=0.48\textwidth]{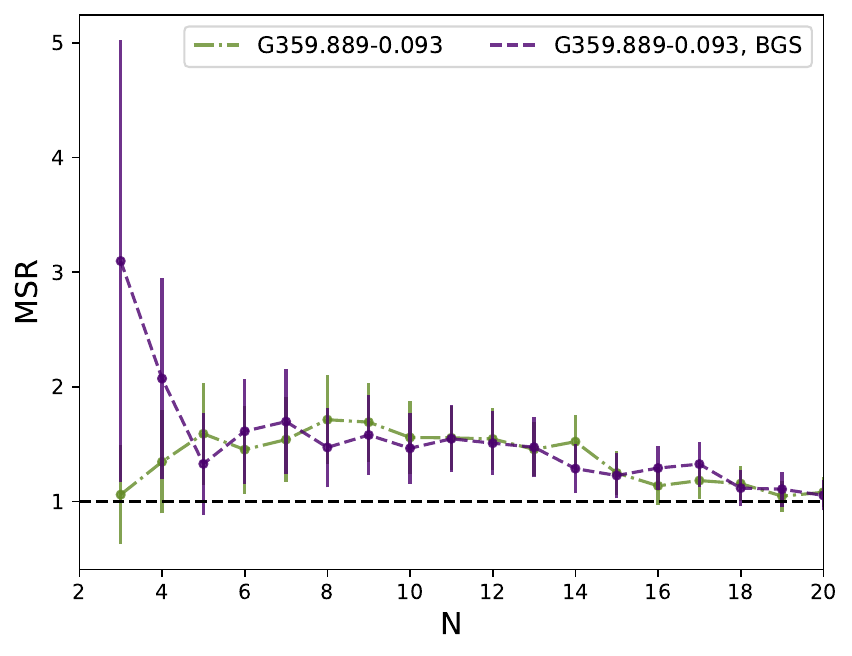}}
\caption{\label{fig.bgscomp} The arithmetic MSR plots for 4 clouds which display the greatest difference between the background-subtracted sources (BGS) and the non-background-subtracted sources. N is limited to N = 20 due to the tendency of the MSR to ``level'' at MSR = 1 for $\frac{N_{total}}{N_{MSR}} \rightarrow 1$. The dotted black line indicates MSR = 1.}
\end{figure*}

As previously stated, we use the non-background-subtracted sources for our calculations in this paper. We calculate the MSRs for the clouds with the background-subtracted sources and observe very few differences between the results. Only four clouds, seen in Figure \ref{fig.bgscomp}, have a distinct difference between the respective arithmetic calculations.

\bibliography{bib}{}

@article{Alcock_2019,
	doi = {10.1093/mnras/stz2646},
	url = {https://doi.org/10.1093%2Fmnras%2Fstz2646},
	year = 2019,
	month = {sep},
	publisher = {Oxford University Press ({OUP})},
	volume = {490},
	number = {1},
	pages = {350--358},
	author = {Hayley L Alcock and Richard J Parker},
	title = {On the mass segregation of cores and stars},
	journal = {Monthly Notices of the Royal Astronomical Society}
}

@article{Allison_2009a,
	doi = {10.1111/j.1365-2966.2009.14508.x},
	url = {https://doi.org/10.1111%2Fj.1365-2966.2009.14508.x},
	year = 2009,
	month = {may},
  	publisher = {Oxford University Press ({OUP})},
  	volume = {395},
  	number = {3},
  	pages = {1449--1454},
  	author = {Richard J. Allison and Simon P. Goodwin and Richard J. Parker and Simon F. Portegies Zwart and Richard de Grijs and M. B. N. Kouwenhoven},
  	title = {Using the minimum spanning tree to trace mass segregation},
  	journal = {Monthly Notices of the Royal Astronomical Society}
}

@article{Barnes_2017,
   title={Star formation rates and efficiencies in the Galactic Centre},
   volume={469},
   ISSN={1365-2966},
   url={http://dx.doi.org/10.1093/mnras/stx941},
   DOI={10.1093/mnras/stx941},
   number={2},
   journal={Monthly Notices of the Royal Astronomical Society},
   publisher={Oxford University Press (OUP)},
   author={Barnes, A. T. and Longmore, S. N. and Battersby, C. and Bally, J. and Kruijssen, J. M. D. and Henshaw, J. D. and Walker, D. L.},
   year={2017},
   month=apr, pages={2263–2285} }

@ARTICLE{Barnes_2019,
       author = {{Barnes}, A.~T. and {Longmore}, S.~N. and {Avison}, A. and {Contreras}, Y. and {Ginsburg}, A. and {Henshaw}, J.~D. and {Rathborne}, J.~M. and {Walker}, D.~L. and {Alves}, J. and {Bally}, J. and {Battersby}, C. and {Beltr{\'a}n}, M.~T. and {Beuther}, H. and {Garay}, G. and {Gomez}, L. and {Jackson}, J. and {Kainulainen}, J. and {Kruijssen}, J.~M.~D. and {Lu}, X. and {Mills}, E.~A.~C. and {Ott}, J. and {Peters}, T.},
        title = "{Young massive star cluster formation in the Galactic Centre is driven by global gravitational collapse of high-mass molecular clouds}",
      journal = {\mnras},
         year = 2019,
        month = jun,
       volume = {486},
       number = {1},
        pages = {283-303},
          doi = {10.1093/mnras/stz796},
archivePrefix = {arXiv},
       eprint = {1903.06158},
 primaryClass = {astro-ph.GA},
       adsurl = {https://ui.adsabs.harvard.edu/abs/2019MNRAS.486..283B}
}

@article{Barrow_1985,
    author = {Barrow, John D. and Bhavsar, Suketu P. and Sonoda, D. H.},
    title = "{Minimal spanning trees, filaments and galaxy clustering}",
    journal = {Monthly Notices of the Royal Astronomical Society},
    volume = {216},
    number = {1},
    pages = {17-35},
    year = {1985},
    month = {09},
    issn = {0035-8711},
    doi = {10.1093/mnras/216.1.17},
    url = {https://doi.org/10.1093/mnras/216.1.17},
    eprint = {https://academic.oup.com/mnras/article-pdf/216/1/17/18521802/mnras216-0017.pdf},
}

@ARTICLE{Longmore_2026,
       author = {{Longmore}, Steven N. and {Bally}, John and {Barnes}, Ashley T. and {Battersby}, Cara and {Colzi}, Laura and {Ginsburg}, Adam and {Henshaw}, Jonathan D. and {Ho}, Paul T.~P. and {Jim{\'e}nez-Serra}, Izaskun and {Kruijssen}, J.~M. Diederik and {Mills}, Elisabeth A.~C. and {Petkova}, Maya A. and {Sormani}, Mattia C. and {Tress}, Robin G. and {Walker}, Daniel L. and {Wallace}, Jennifer and {Alkhuja}, Emad and {Armillotta}, Lucia and {Budaiev}, Nazar and {Buddhacharya}, Rojita and {Bulatek}, Alyssa and {Burton}, Michael and {Butterfield}, Natalie O. and {Busch}, Laura A. and {Caselli}, Paola and {Chevance}, M{\'e}lanie and {Cook}, Claire and {Crowe}, Samuel and {D{\'\i}az-Rodr{\'\i}guez}, Ana Karla and {DiTeodoro}, Enrico and {Dicker}, Simon R. and {Dutkowska}, Katarzyna M. and {Fairley}, Adam and {Federrath}, Christoph and {Fedriani}, Rub{\'e}n and {Feng}, Zi-Xuan and {Fiteni}, Karl and {Fuller}, Gary and {Garc{\'\i}a}, Pablo and {Goicoechea}, Javier and {Girichidis}, Philipp and {Glover}, Simon C.~O. and {Gorski}, Mark and {Gramze}, Savannah R. and {Gu}, Qi-Lao and {Hatchfield}, H. Perry and {Henkel}, Christian and {Houghton}, Rebecca J. and {Hsieh}, Pei-Ying and {Hu}, Yue and {Immer}, Katharina and {Jeff}, Desmond and {Karoly}, Janik and {Kauffmann}, Jens and {Klessen}, Ralf S. and {Krumholz}, Mark R. and {Lazarian}, Alex and {Levesque}, Emily M. and {Liang}, Fu-Heng and {Lipman}, Dani and {Liu}, Xunchuan and {Lu}, Xing and {Luo}, Qiu-yi and {Lupi}, Alessandro and {McCafferty}, Laura and {Mart{\'\i}n}, S. and {Mazoochi}, Farideh and {Morris}, Mark R. and {Nonhebel}, Marie and {Nogueras-Lara}, Francisco and {Oka}, Tomoharu and {Ott}, Juergen and {Padovani}, Marco and {Pan}, Xing and {Pineda}, Jaime E. and {Pillai}, Thushara G.~S. and {Pound}, Marc W. and {Requena Torres}, Miguel and {Riquelme-V{\'a}squez}, Denise and {Rivilla}, V{\'\i}ctor M. and {Salo}, Galaxy and {S{\'a}nchez-Monge}, {\'A}lvaro and {Santa-Maria}, Miriam G. and {Schoedel}, Rainer and {Schmiedeke}, Anika and {Schultheis}, Matthias and {Smith}, Howard A. and {Sofue}, Yoshiaki and {Testi}, Leonardo and {Tremblay}, Grant R. and {Vasini}, Arianna and {Vermari{\"e}n}, Gijs and {Vikhlinin}, Alexey and {Viti}, Serena and {Wang}, Q. Daniel and {Xu}, Fengwei and {Zhang}, Suinan and {Zhang}, Qizhou},
        title = "{ALMA Central Molecular Zone Exploration Survey (ACES) I: Overview}",
      journal = {arXiv e-prints},
         year = 2026,
        month = feb,
          eid = {arXiv:2602.20340},
        pages = {arXiv:2602.20340},
          doi = {10.48550/arXiv.2602.20340},
archivePrefix = {arXiv},
       eprint = {2602.20340},
 primaryClass = {astro-ph.GA},
       adsurl = {https://ui.adsabs.harvard.edu/abs/2026arXiv260220340L}
}

@article{Battersby_2011,
   title={Characterizing precursors to stellar clusters withHerschel},
   volume={535},
   ISSN={1432-0746},
   url={http://dx.doi.org/10.1051/0004-6361/201116559},
   DOI={10.1051/0004-6361/201116559},
   journal={Astronomy \& Astrophysics},
   publisher={EDP Sciences},
   author={Battersby, C. and Bally, J. and Ginsburg, A. and Bernard, J.-P. and Brunt, C. and Fuller, G. A. and Martin, P. and Molinari, S. and Mottram, J. and Peretto, N. and Testi, L. and Thompson, M. A.},
   year={2011},
   month=nov, pages={A128} }

@ARTICLE{Battersby_2020,
       author = {{Battersby}, Cara and {Keto}, Eric and {Walker}, Daniel and {Barnes}, Ashley and {Callanan}, Daniel and {Ginsburg}, Adam and {Hatchfield}, H. Perry and {Henshaw}, Jonathan and {Kauffmann}, Jens and {Kruijssen}, J.~M. Diederik and {Longmore}, Steven N. and {Lu}, Xing and {Mills}, Elisabeth A.~C. and {Pillai}, Thushara and {Zhang}, Qizhou and {Bally}, John and {Butterfield}, Natalie and {Contreras}, Yanett A. and {Ho}, Luis C. and {Ott}, J{\"u}rgen and {Patel}, Nimesh and {Tolls}, Volker},
        title = "{CMZoom: Survey Overview and First Data Release}",
      journal = {\apjs},
         year = 2020,
        month = aug,
       volume = {249},
       number = {2},
          eid = {35},
        pages = {35},
          doi = {10.3847/1538-4365/aba18e},
archivePrefix = {arXiv},
       eprint = {2007.05023},
 primaryClass = {astro-ph.GA},
       adsurl = {https://ui.adsabs.harvard.edu/abs/2020ApJS..249...35B}
}

@misc{Battersby_2024a,
      title={3-D CMZ I: Central Molecular Zone Overview}, 
      author={Cara Battersby and Daniel L. Walker and Ashley Barnes and Adam Ginsburg and Dani Lipman and Danya Alboslani and H Perry Hatchfield and John Bally and Simon C. O. Glover and Jonathan D. Henshaw and Katharina Immer and Ralf S. Klessen and Steven N. Longmore and Elisabeth A. C. Mills and Sergio Molinari and Rowan Smith and Mattia C. Sormani and Robin G. Tress and Qizhou Zhang},
      year={2024},
      eprint={2410.17334},
      archivePrefix={arXiv},
      primaryClass={astro-ph.GA},
      url={https://arxiv.org/abs/2410.17334}, 
}

@ARTICLE{Battersby_2025a,
       author = {{Battersby}, Cara and {Walker}, Daniel L. and {Barnes}, Ashley and {Ginsburg}, Adam and {Lipman}, Dani and {Alboslani}, Danya and {Hatchfield}, H. Perry and {Bally}, John and {Glover}, Simon C.~O. and {Henshaw}, Jonathan D. and {Immer}, Katharina and {Klessen}, Ralf S. and {Longmore}, Steven N. and {Mills}, Elisabeth A.~C. and {Molinari}, Sergio and {Smith}, Rowan and {Sormani}, Mattia C. and {Tress}, Robin G. and {Zhang}, Qizhou},
        title = "{3D CMZ. I. Central Molecular Zone Overview}",
      journal = {\apj},
         year = 2025,
        month = may,
       volume = {984},
       number = {2},
          eid = {156},
        pages = {156},
          doi = {10.3847/1538-4357/adb5f0},
archivePrefix = {arXiv},
       eprint = {2410.17334},
 primaryClass = {astro-ph.GA},
       adsurl = {https://ui.adsabs.harvard.edu/abs/2025ApJ...984..156B}
}

@ARTICLE{Battersby_2025b,
       author = {{Battersby}, Cara and {Walker}, Daniel L. and {Barnes}, Ashley and {Ginsburg}, Adam and {Lipman}, Dani and {Alboslani}, Danya and {Hatchfield}, H. Perry and {Bally}, John and {Glover}, Simon C.~O. and {Henshaw}, Jonathan D. and {Immer}, Katharina and {Klessen}, Ralf S. and {Longmore}, Steven N. and {Mills}, Elisabeth A.~C. and {Molinari}, Sergio and {Smith}, Rowan and {Sormani}, Mattia C. and {Tress}, Robin G. and {Zhang}, Qizhou},
        title = "{3D CMZ. II. Hierarchical Structure Analysis of the Central Molecular Zone}",
      journal = {\apj},
         year = 2025,
        month = may,
       volume = {984},
       number = {2},
          eid = {157},
        pages = {157},
          doi = {10.3847/1538-4357/adb844},
archivePrefix = {arXiv},
       eprint = {2410.17332},
 primaryClass = {astro-ph.GA},
       adsurl = {https://ui.adsabs.harvard.edu/abs/2025ApJ...984..157B}
}

@ARTICLE{Beuther_2012,
       author = {{Beuther}, H. and {Tackenberg}, J. and {Linz}, H. and {Henning}, Th. and {Schuller}, F. and {Wyrowski}, F. and {Schilke}, P. and {Menten}, K. and {Robitaille}, T.~P. and {Walmsley}, C.~M. and {Bronfman}, L. and {Motte}, F. and {Nguyen-Luong}, Q. and {Bontemps}, S.},
        title = "{Galactic Structure Based on the ATLASGAL 870 {\ensuremath{\mu}}m Survey}",
      journal = {\apj},
         year = 2012,
        month = mar,
       volume = {747},
       number = {1},
          eid = {43},
        pages = {43},
          doi = {10.1088/0004-637X/747/1/43},
archivePrefix = {arXiv},
       eprint = {1112.4609},
 primaryClass = {astro-ph.SR},
       adsurl = {https://ui.adsabs.harvard.edu/abs/2012ApJ...747...43B}
}

@ARTICLE{Callanan_2023,
       author = {{Callanan}, Daniel and {Longmore}, Steven N. and {Battersby}, Cara and {Hatchfield}, H. Perry and {Walker}, Daniel L. and {Henshaw}, Jonathan and {Keto}, Eric and {Barnes}, Ashley and {Ginsburg}, Adam and {Kauffmann}, Jens and {Kruijssen}, J.~M. Diederik and {Lu}, Xing and {Mills}, Elisabeth A.~C. and {Pillai}, Thushara and {Zhang}, Qizhou and {Bally}, John and {Butterfield}, Natalie and {Contreras}, Yanett A. and {Ho}, Luis C. and {Immer}, Katharina and {Johnston}, Katharine G. and {Ott}, Juergen and {Patel}, Nimesh and {Tolls}, Volker},
        title = "{CMZoom III: Spectral line data release}",
      journal = {\mnras},
         year = 2023,
        month = apr,
       volume = {520},
       number = {3},
        pages = {4760-4778},
          doi = {10.1093/mnras/stad388},
archivePrefix = {arXiv},
       eprint = {2301.04699},
 primaryClass = {astro-ph.GA},
       adsurl = {https://ui.adsabs.harvard.edu/abs/2023MNRAS.520.4760C}
}

@ARTICLE{Dahmen_1998,
       author = {{Dahmen}, G. and {Huttemeister}, S. and {Wilson}, T.~L. and {Mauersberger}, R.},
        title = "{Molecular gas in the Galactic center region. II. Gas mass and N\_, = H\_2/I\_\^(12)CO conversion based on a C\^(18)O(J = 1 -> 0) survey}",
      journal = {\aap},
         year = 1998,
        month = mar,
       volume = {331},
        pages = {959-976},
          doi = {10.48550/arXiv.astro-ph/9711117},
archivePrefix = {arXiv},
       eprint = {astro-ph/9711117},
 primaryClass = {astro-ph},
       adsurl = {https://ui.adsabs.harvard.edu/abs/1998A&A...331..959D}
}

@article{Dib_2019,
	doi = {10.1051/0004-6361/201834080},
	url = {https://doi.org/10.1051%2F0004-6361%2F201834080},
	year = 2019,
	month = {sep},
	publisher = {{EDP} Sciences},
	volume = {629},
	pages = {A135},
	author = {Sami Dib and Thomas Henning},
	title = {Star formation activity and the spatial distribution and mass segregation of dense cores in the early phases of star formation},
	journal = {Astronomy \& Astrophysics}
}

@ARTICLE{Dib_2018,
       author = {{Dib}, Sami and {Schmeja}, Stefan and {Parker}, Richard J.},
        title = "{Structure and mass segregation in Galactic stellar clusters}",
      journal = {\mnras},
         year = 2018,
        month = jan,
       volume = {473},
       number = {1},
        pages = {849-859},
          doi = {10.1093/mnras/stx2413},
archivePrefix = {arXiv},
       eprint = {1707.00744},
 primaryClass = {astro-ph.GA},
       adsurl = {https://ui.adsabs.harvard.edu/abs/2018MNRAS.473..849D}
}

@article{Federrath_2016,
   title={THE LINK BETWEEN TURBULENCE, MAGNETIC FIELDS, FILAMENTS, AND STAR FORMATION IN THE CENTRAL MOLECULAR ZONE CLOUD G0.253+0.016},
   volume={832},
   ISSN={1538-4357},
   url={http://dx.doi.org/10.3847/0004-637X/832/2/143},
   DOI={10.3847/0004-637x/832/2/143},
   number={2},
   journal={The Astrophysical Journal},
   publisher={American Astronomical Society},
   author={Federrath, C. and Rathborne, J. M. and Longmore, S. N. and Kruijssen, J. M. D. and Bally, J. and Contreras, Y. and Crocker, R. M. and Garay, G. and Jackson, J. M. and Testi, L. and Walsh, A. J.},
   year={2016},
   month=nov, pages={143} }

@article{Ginsburg_2016,
   title={Dense gas in the Galactic central molecular zone is warm and heated by turbulence},
   volume={586},
   ISSN={1432-0746},
   url={http://dx.doi.org/10.1051/0004-6361/201526100},
   DOI={10.1051/0004-6361/201526100},
   journal={Astronomy \& Astrophysics},
   publisher={EDP Sciences},
   author={Ginsburg, Adam and Henkel, Christian and Ao, Yiping and Riquelme, Denise and Kauffmann, Jens and Pillai, Thushara and Mills, Elisabeth A. C. and Requena-Torres, Miguel A. and Immer, Katharina and Testi, Leonardo and Ott, Juergen and Bally, John and Battersby, Cara and Darling, Jeremy and Aalto, Susanne and Stanke, Thomas and Kendrew, Sarah and Diederik Kruijssen, J. M. and Longmore, Steven and Dale, James and Guesten, Rolf and Menten, Karl M.},
   year={2016},
   month=jan, pages={A50} }

@article{Ginsburg_2018a,
   title={Distributed Star Formation throughout the Galactic Center Cloud Sgr B2},
   volume={853},
   ISSN={1538-4357},
   url={http://dx.doi.org/10.3847/1538-4357/aaa6d4},
   DOI={10.3847/1538-4357/aaa6d4},
   number={2},
   journal={The Astrophysical Journal},
   publisher={American Astronomical Society},
   author={Ginsburg, Adam and Bally, John and Barnes, Ashley and Bastian, Nate and Battersby, Cara and Beuther, Henrik and Brogan, Crystal and Contreras, Yanett and Corby, Joanna and Darling, Jeremy and De Pree, Chris and Galván-Madrid, Roberto and Garay, Guido and Henshaw, Jonathan and Hunter, Todd and Kruijssen, J. M. Diederik and Longmore, Steven and Lu, Xing and Meng, Fanyi and Mills, Elisabeth A. C. and Ott, Juergen and Pineda, Jaime E. and Sánchez-Monge, Álvaro and Schilke, Peter and Schmiedeke, Anika and Walker, Daniel and Wilner, David},
   year={2018},
   month=feb, pages={171} }

@ARTICLE{Guesten_1983,
       author = {{Guesten}, R. and {Downes}, D.},
        title = "{New H2O masers in the galactic center region.}",
      journal = {\aap},
         year = 1983,
        month = jan,
       volume = {117},
        pages = {343-346},
       adsurl = {https://ui.adsabs.harvard.edu/abs/1983A&A...117..343G}
}

@article{Hatchfield_2020,
	doi = {10.3847/1538-4365/abb610},
	url = {https://doi.org/10.3847%2F1538-4365%2Fabb610},
	year = 2020,
	month = {nov},
	publisher = {American Astronomical Society},
	volume = {251},
	number = {1},
	pages = {14},
	author = {H Perry Hatchfield and Cara Battersby and Eric Keto and Daniel Walker and Ashley Barnes and Daniel Callanan and Adam Ginsburg and Jonathan D. Henshaw and Jens Kauffmann and J. M. Diederik Kruijssen and Steve N. Longmore and Xing Lu and Elisabeth A. C. Mills and Thushara Pillai and Qizhou Zhang and John Bally and Natalie Butterfield and Yanett A. Contreras and Luis C. Ho and Jürgen Ott and Nimesh Patel and Volker Tolls},
	title = {{CMZoom}. {II}. Catalog of Compact Submillimeter Dust Continuum Sources in the Milky Way's Central Molecular Zone},
	journal = {The Astrophysical Journal Supplement Series}
}

@ARTICLE{Hatchfield_2024,
       author = {{Hatchfield}, H. Perry and {Battersby}, Cara and {Barnes}, Ashley T. and {Butterfield}, Natalie and {Ginsburg}, Adam and {Henshaw}, Jonathan D. and {Longmore}, Steven N. and {Lu}, Xing and {Svoboda}, Brian and {Walker}, Daniel and {Callanan}, Daniel and {Mills}, Elisabeth A.~C. and {Ho}, Luis C. and {Kauffmann}, Jens and {Kruijssen}, J.~M. Diederik and {Ott}, J{\"u}rgen and {Pillai}, Thushara and {Zhang}, Qizhou},
        title = "{CMZoom. IV. Incipient High-mass Star Formation throughout the Central Molecular Zone}",
      journal = {\apj},
         year = 2024,
        month = feb,
       volume = {962},
       number = {1},
          eid = {14},
        pages = {14},
          doi = {10.3847/1538-4357/ad10af},
archivePrefix = {arXiv},
       eprint = {2312.09284},
 primaryClass = {astro-ph.GA},
       adsurl = {https://ui.adsabs.harvard.edu/abs/2024ApJ...962...14H}
}

@misc{Henshaw_2022,
      title={Star Formation in the Central Molecular Zone of the Milky Way}, 
      author={Jonathan D. Henshaw and Ashley T. Barnes and Cara Battersby and Adam Ginsburg and Mattia C. Sormani and Daniel L. Walker},
      year={2022},
      eprint={2203.11223},
      archivePrefix={arXiv},
      primaryClass={astro-ph.GA},
      url={https://arxiv.org/abs/2203.11223}, 
}

@article{Immer_2012,
   title={Recent star formation in the inner Galactic bulge seen by ISOGAL: II. The central molecular zone},
   volume={537},
   ISSN={1432-0746},
   url={http://dx.doi.org/10.1051/0004-6361/201117857},
   DOI={10.1051/0004-6361/201117857},
   journal={Astronomy \& Astrophysics},
   publisher={EDP Sciences},
   author={Immer, K. and Schuller, F. and Omont, A. and Menten, K. M.},
   year={2012},
   month=jan, pages={A121} }

@article{Jones_2013,
   title={Spectral imaging of the central molecular zone in multiple 7-mm molecular lines},
   volume={433},
   ISSN={1365-2966},
   url={http://dx.doi.org/10.1093/mnras/stt717},
   DOI={10.1093/mnras/stt717},
   number={1},
   journal={Monthly Notices of the Royal Astronomical Society},
   publisher={Oxford University Press (OUP)},
   author={Jones, P. A. and Burton, M. G. and Cunningham, M. R. and Tothill, N. F. H. and Walsh, A. J.},
   year={2013},
   month=may, pages={221–234} }

@ARTICLE{Kauffmann_2008,
       author = {{Kauffmann}, J. and {Bertoldi}, F. and {Bourke}, T.~L. and {Evans}, N.~J., II and {Lee}, C.~W.},
        title = "{MAMBO mapping of Spitzer c2d small clouds and cores}",
      journal = {\aap},
         year = 2008,
        month = sep,
       volume = {487},
       number = {3},
        pages = {993-1017},
          doi = {10.1051/0004-6361:200809481},
archivePrefix = {arXiv},
       eprint = {0805.4205},
 primaryClass = {astro-ph},
       adsurl = {https://ui.adsabs.harvard.edu/abs/2008A&A...487..993K}
}

@article{Kauffmann_2017b,
   title={The Galactic Center Molecular Cloud Survey: I. A steep linewidth-size relation and suppression of star formation},
   volume={603},
   ISSN={1432-0746},
   url={http://dx.doi.org/10.1051/0004-6361/201628088},
   DOI={10.1051/0004-6361/201628088},
   journal={Astronomy \& Astrophysics},
   publisher={EDP Sciences},
   author={Kauffmann, Jens and Pillai, Thushara and Zhang, Qizhou and Menten, Karl M. and Goldsmith, Paul F. and Lu, Xing and Guzmán, Andrés E.},
   year={2017},
   month=jul, pages={A89} }

@ARTICLE{Kauffmann_2017a,
       author = {{Kauffmann}, Jens and {Pillai}, Thushara and {Zhang}, Qizhou and {Menten}, Karl M. and {Goldsmith}, Paul F. and {Lu}, Xing and {Guzm{\'a}n}, Andr{\'e}s E. and {Schmiedeke}, Anika},
        title = "{The Galactic Center Molecular Cloud Survey. II. A lack of dense gas and cloud evolution along Galactic center orbits}",
      journal = {\aap},
         year = 2017,
        month = jul,
       volume = {603},
          eid = {A90},
        pages = {A90},
          doi = {10.1051/0004-6361/201628089},
archivePrefix = {arXiv},
       eprint = {1610.03502},
 primaryClass = {astro-ph.GA},
       adsurl = {https://ui.adsabs.harvard.edu/abs/2017A&A...603A..90K}
}

@article{Kinman_2024,
      title={The Core Mass Function Across Galactic Environments. IV. The Galactic Center}, 
      author={Alva V. I. Kinman and Maya A. Petkova and Jonathan C. Tan and Giuliana Cosentino and Yu Cheng},
      year={2024},
      eprint={2403.04032},
      archivePrefix={arXiv},
      primaryClass={astro-ph.GA}
}

@ARTICLE{Kruijssen_2019,
       author = {{Kruijssen}, J.~M.~D. and {Dale}, J.~E. and {Longmore}, S.~N. and {Walker}, D.~L. and {Henshaw}, J.~D. and {Jeffreson}, S.~M.~R. and {Petkova}, M.~A. and {Ginsburg}, A. and {Barnes}, A.~T. and {Battersby}, C.~D. and {Immer}, K. and {Jackson}, J.~M. and {Keto}, E.~R. and {Krieger}, N. and {Mills}, E.~A.~C. and {S{\'a}nchez-Monge}, {\'A}. and {Schmiedeke}, A. and {Suri}, S.~T. and {Zhang}, Q.},
        title = "{The dynamical evolution of molecular clouds near the Galactic Centre - II. Spatial structure and kinematics of simulated clouds}",
      journal = {\mnras},
         year = 2019,
        month = apr,
       volume = {484},
       number = {4},
        pages = {5734-5754},
          doi = {10.1093/mnras/stz381},
archivePrefix = {arXiv},
       eprint = {1902.01860},
 primaryClass = {astro-ph.GA},
       adsurl = {https://ui.adsabs.harvard.edu/abs/2019MNRAS.484.5734K}
}

@article{Lu_2019a,
   title={A Census of Early-phase High-mass Star Formation in the Central Molecular Zone},
   volume={244},
   ISSN={1538-4365},
   url={http://dx.doi.org/10.3847/1538-4365/ab4258},
   DOI={10.3847/1538-4365/ab4258},
   number={2},
   journal={The Astrophysical Journal Supplement Series},
   publisher={American Astronomical Society},
   author={Lu, Xing and Mills, Elisabeth A. C. and Ginsburg, Adam and Walker, Daniel L. and Barnes, Ashley T. and Butterfield, Natalie and Henshaw, Jonathan D. and Battersby, Cara and Kruijssen, J. M. Diederik and Longmore, Steven N. and Zhang, Qizhou and Bally, John and Kauffmann, Jens and Ott, Jürgen and Rickert, Matthew and Wang, Ke},
   year={2019},
   month=oct, pages={35} }

@ARTICLE{Lu_2020,
       author = {{Lu}, Xing and {Cheng}, Yu and {Ginsburg}, Adam and {Longmore}, Steven N. and {Kruijssen}, J.~M. Diederik and {Battersby}, Cara and {Zhang}, Qizhou and {Walker}, Daniel L.},
        title = "{ALMA Observations of Massive Clouds in the Central Molecular Zone: Jeans Fragmentation and Cluster Formation}",
      journal = {\apjl},
         year = 2020,
        month = may,
       volume = {894},
       number = {2},
          eid = {L14},
        pages = {L14},
          doi = {10.3847/2041-8213/ab8b65},
archivePrefix = {arXiv},
       eprint = {2004.09532},
 primaryClass = {astro-ph.GA},
       adsurl = {https://ui.adsabs.harvard.edu/abs/2020ApJ...894L..14L}
}

@ARTICLE{Lu_2024,
       author = {{Lu}, Xing and {Liu}, Junhao and {Pillai}, Thushara and {Zhang}, Qizhou and {Liu}, Tie and {Gu}, Qilao and {Hasegawa}, Tetsuo and {Li}, Pak Shing and {Tang}, Xindi and {Hatchfield}, H. Perry and {Issac}, Namitha and {Liu}, Xunchuan and {Luo}, Qiuyi and {Mai}, Xiaofeng and {Shen}, Zhiqiang},
        title = "{Magnetic Fields in the Central Molecular Zone Influenced by Feedback and Weakly Correlated with Star Formation}",
      journal = {\apj},
         year = 2024,
        month = feb,
       volume = {962},
       number = {1},
          eid = {39},
        pages = {39},
          doi = {10.3847/1538-4357/ad1395},
archivePrefix = {arXiv},
       eprint = {2312.01776},
 primaryClass = {astro-ph.GA},
       adsurl = {https://ui.adsabs.harvard.edu/abs/2024ApJ...962...39L}
}

@article{Mac_Low_2004,
   title={Control of star formation by supersonic turbulence},
   volume={76},
   ISSN={1539-0756},
   url={http://dx.doi.org/10.1103/RevModPhys.76.125},
   DOI={10.1103/revmodphys.76.125},
   number={1},
   journal={Reviews of Modern Physics},
   publisher={American Physical Society (APS)},
   author={Mac Low, Mordecai-Mark and Klessen, Ralf S.},
   year={2004},
   month=jan, pages={125–194} }

@article{Mangilli_2019,
   title={The geometry of the magnetic field in the central molecular zone measured by PILOT},
   volume={630},
   ISSN={1432-0746},
   url={http://dx.doi.org/10.1051/0004-6361/201935072},
   DOI={10.1051/0004-6361/201935072},
   journal={Astronomy \& Astrophysics},
   publisher={EDP Sciences},
   author={Mangilli, A. and Aumont, J. and Bernard, J.-Ph. and Buzzelli, A. and de Gasperis, G. and Durrive, J. B. and Ferriere, K. and Foënard, G. and Hughes, A. and Lacourt, A. and Misawa, R. and Montier, L. and Mot, B. and Ristorcelli, I. and Roussel, H. and Ade, P. and Alina, D. and de Bernardis, P. and de Gouveia Dal Pino, E. and Dubois, J. P. and Engel, C. and Guillet, V. and Hargrave, P. and Laureijs, R. and Longval, Y. and Maffei, B. and Magalhes, A. M. and Marty, C. and Masi, S. and Montel, J. and Pajot, F. and Pérot, E. and Rodriguez, L. and Salatino, M. and Saccoccio, M. and Savini, G. and Stever, S. and Tauber, J. and Tibbs, C. and Tucker, C.},
   year={2019},
   month=sep, pages={A74} }

@ARTICLE{Maschberger_2011,
       author = {{Maschberger}, Th. and {Clarke}, C.~J.},
        title = "{Global mass segregation in hydrodynamical simulations of star formation}",
      journal = {\mnras},
         year = 2011,
        month = sep,
       volume = {416},
       number = {1},
        pages = {541-546},
          doi = {10.1111/j.1365-2966.2011.19067.x},
archivePrefix = {arXiv},
       eprint = {1106.1181},
 primaryClass = {astro-ph.GA},
       adsurl = {https://ui.adsabs.harvard.edu/abs/2011MNRAS.416..541M}
}

@article{Mills_2013,
   title={DETECTION OF WIDESPREAD HOT AMMONIA IN THE GALACTIC CENTER},
   volume={772},
   ISSN={1538-4357},
   url={http://dx.doi.org/10.1088/0004-637X/772/2/105},
   DOI={10.1088/0004-637x/772/2/105},
   number={2},
   journal={The Astrophysical Journal},
   publisher={American Astronomical Society},
   author={Mills, E. A. C. and Morris, M. R.},
   year={2013},
   month=jul, pages={105} }

@article{Mills_2015,
   title={ABUNDANT CH3OH MASERS BUT NO NEW EVIDENCE FOR STAR FORMATION IN GCM0.253+0.016},
   volume={805},
   ISSN={1538-4357},
   url={http://dx.doi.org/10.1088/0004-637X/805/1/72},
   DOI={10.1088/0004-637x/805/1/72},
   number={1},
   journal={The Astrophysical Journal},
   publisher={American Astronomical Society},
   author={Mills, E. A. C. and Butterfield, N. and Ludovici, D. A. and Lang, C. C. and Ott, J. and Morris, M. R. and Schmitz, S.},
   year={2015},
   month=may, pages={72} }

@ARTICLE{Mills_2018,
       author = {{Mills}, E.~A.~C. and {Ginsburg}, A. and {Immer}, K. and {Barnes}, J.~M. and {Wiesenfeld}, L. and {Faure}, A. and {Morris}, M.~R. and {Requena-Torres}, M.~A.},
        title = "{The Dense Gas Fraction in Galactic Center Clouds}",
      journal = {\apj},
         year = 2018,
        month = nov,
       volume = {868},
       number = {1},
          eid = {7},
        pages = {7},
          doi = {10.3847/1538-4357/aae581},
archivePrefix = {arXiv},
       eprint = {1810.00266},
 primaryClass = {astro-ph.GA},
       adsurl = {https://ui.adsabs.harvard.edu/abs/2018ApJ...868....7M}
}

@article{Morii_2023,
author = {Morii, Kaho and Sanhueza, Patricio and Nakamura, Fumitaka and Zhang, Qizhou and Sabatini, Giovanni and Beuther, Henrik and Lu, An and Li, Shanghuo and Garay, Guido and Jackson, James and Olguin, Fernando and Tafoya, D. and Tatematsu, Ken'ichi and Izumi, Natsuko and Sakai, Takeshi and Silva, Andrea},
year = {2023},
month = {06},
pages = {148},
title = {The ALMA Survey of 70 μm Dark High-mass Clumps in Early Stages (ASHES). IX. Physical Properties and Spatial Distribution of Cores in IRDCs},
volume = {950},
journal = {The Astrophysical Journal},
doi = {10.3847/1538-4357/acccea}
}

@article{Nony_2021,
	doi = {10.1051/0004-6361/202039353},
  
	url = {https://doi.org/10.1051%2F0004-6361%2F202039353},
  
	year = 2021,
	month = {jan},
  
	publisher = {{EDP} Sciences},
  
	volume = {645},
  
	pages = {A94},
  
	author = {T. Nony and J.-F. Robitaille and F. Motte and M. Gonzalez and I. Joncour and E. Moraux and A. Men'shchikov and P. Didelon and F. Louvet and A. S. M. Buckner and N. Schneider and S. L. Lumsden and S. Bontemps and Y. Pouteau and N. Cunningham and E. Fiorellino and R. Oudmaijer and P. Andr{\'{e}
} and B. Thomasson},
  
	title = {Mass segregation and sequential star formation in {NGC} 2264 revealed by $\less$i$\greater$Herschel$\less$/i$\greater$},
  
	journal = {Astronomy \& Astrophysics}
}

@ARTICLE{Paglione_1998,
       author = {{Paglione}, Timothy A.~D. and {Jackson}, James M. and {Bolatto}, Alberto D. and {Heyer}, Mark H.},
        title = "{Interpreting the HCN/CO Intensity Ratio in the Galactic Center}",
      journal = {\apj},
         year = 1998,
        month = jan,
       volume = {493},
       number = {2},
        pages = {680-693},
          doi = {10.1086/305136},
       adsurl = {https://ui.adsabs.harvard.edu/abs/1998ApJ...493..680P}
}

@article{Pavl_k_2019,
   title={Do star clusters form in a completely mass-segregated way?},
   volume={626},
   ISSN={1432-0746},
   url={http://dx.doi.org/10.1051/0004-6361/201834265},
   DOI={10.1051/0004-6361/201834265},
   journal={Astronomy \& Astrophysics},
   publisher={EDP Sciences},
   author={Pavlík, Václav and Kroupa, Pavel and Šubr, Ladislav},
   year={2019},
   month=jun, pages={A79} }

@article{Pokhrel_2018,
doi = {10.3847/1538-4357/aaa240},
url = {https://doi.org/10.3847/1538-4357/aaa240},
year = {2018},
month = {jan},
publisher = {The American Astronomical Society},
volume = {853},
number = {1},
pages = {5},
author = {Pokhrel, Riwaj and Myers, Philip C. and Dunham, Michael M. and Stephens, Ian W. and Sadavoy, Sarah I. and Zhang, Qizhou and Bourke, Tyler L. and Tobin, John J. and Lee, Katherine I. and Gutermuth, Robert A. and Offner, Stella S. R.},
title = {Hierarchical Fragmentation in the Perseus Molecular Cloud: From the Cloud Scale to Protostellar Objects},
journal = {The Astrophysical Journal}
}

@ARTICLE{Pouteau_2023,
       author = {{Pouteau}, Y. and {Motte}, F. and {Nony}, T. and {Gonz{\'a}lez}, M. and {Joncour}, I. and {Robitaille}, J. -F. and {Busquet}, G. and {Galv{\'a}n-Madrid}, R. and {Gusdorf}, A. and {Hennebelle}, P. and {Ginsburg}, A. and {Csengeri}, T. and {Sanhueza}, P. and {Dell'Ova}, P. and {Stutz}, A.~M. and {Towner}, A.~P.~M. and {Cunningham}, N. and {Louvet}, F. and {Men'shchikov}, A. and {Fern{\'a}ndez-L{\'o}pez}, M. and {Schneider}, N. and {Armante}, M. and {Bally}, J. and {Baug}, T. and {Bonfand}, M. and {Bontemps}, S. and {Bronfman}, L. and {Brouillet}, N. and {D{\'\i}az-Gonz{\'a}lez}, D. and {Herpin}, F. and {Lefloch}, B. and {Liu}, H. -L. and {Lu}, X. and {Nakamura}, F. and {Nguyen-Luong}, Q. and {Olguin}, F. and {Tatematsu}, K. and {Valeille-Manet}, M.},
        title = "{ALMA-IMF. VI. Investigating the origin of stellar masses: Core mass function evolution in the W43-MM2\&MM3 mini-starburst}",
      journal = {\aap},
         year = 2023,
        month = jun,
       volume = {674},
          eid = {A76},
        pages = {A76},
          doi = {10.1051/0004-6361/202244776},
archivePrefix = {arXiv},
       eprint = {2212.09307},
 primaryClass = {astro-ph.GA},
       adsurl = {https://ui.adsabs.harvard.edu/abs/2023A&A...674A..76P}
}

@article{Olczak_2011a,
      title={Dynamics in Young Star Clusters: From Planets to Massive Stars}, 
      author={C. Olczak and R. Spurzem and Th. Henning and T. Kaczmarek and S. Pfalzner and S. Harfst and S. Portegies Zwart},
      year={2011},
      eprint={1108.2446},
      archivePrefix={arXiv},
      primaryClass={astro-ph.GA}
}

@article{Parker_2015,
	doi = {10.1093/mnras/stv539},
  
	url = {https://doi.org/10.1093%2Fmnras%2Fstv539},
  
	year = 2015,
	month = {apr},
  
	publisher = {Oxford University Press ({OUP})},
  
	volume = {449},
  
	number = {4},
  
	pages = {3381--3392},
  
	author = {Richard J. Parker and Simon P. Goodwin},
  
	title = {Comparisons between different techniques for measuring mass segregation},
  
	journal = {Monthly Notices of the Royal Astronomical Society}
}

@article{Parker_2018,
	doi = {10.1093/mnras/sty249},
	url = {https://doi.org/10.1093%2Fmnras%2Fsty249},
	year = 2018,
	month = {jan},
	publisher = {Oxford University Press ({OUP})},
	volume = {476},
	number = {1},
	pages = {617--629},
	author = {Richard J Parker},
	title = {On the spatial distributions of dense cores in Orion B},
	journal = {Monthly Notices of the Royal Astronomical Society}
}

@ARTICLE{Prim_1957,
  author={Prim, R. C.},
  journal={The Bell System Technical Journal}, 
  title={Shortest connection networks and some generalizations}, 
  year={1957},
  volume={36},
  number={6},
  pages={1389-1401},
  doi={10.1002/j.1538-7305.1957.tb01515.x}}

@ARTICLE{Sanhueza_2019,
       author = {{Sanhueza}, Patricio and {Contreras}, Yanett and {Wu}, Benjamin and {Jackson}, James M. and {Guzm{\'a}n}, Andr{\'e}s E. and {Zhang}, Qizhou and {Li}, Shanghuo and {Lu}, Xing and {Silva}, Andrea and {Izumi}, Natsuko and {Liu}, Tie and {Miura}, Rie E. and {Tatematsu}, Ken'ichi and {Sakai}, Takeshi and {Beuther}, Henrik and {Garay}, Guido and {Ohashi}, Satoshi and {Saito}, Masao and {Nakamura}, Fumitaka and {Saigo}, Kazuya and {Veena}, V.~S. and {Nguyen-Luong}, Quang and {Tafoya}, Daniel},
        title = "{The ALMA Survey of 70 {\ensuremath{\mu}}m Dark High-mass Clumps in Early Stages (ASHES). I. Pilot Survey: Clump Fragmentation}",
      journal = {\apj},
         year = 2019,
        month = dec,
       volume = {886},
       number = {2},
          eid = {102},
        pages = {102},
          doi = {10.3847/1538-4357/ab45e9},
archivePrefix = {arXiv},
       eprint = {1909.07985},
 primaryClass = {astro-ph.GA},
       adsurl = {https://ui.adsabs.harvard.edu/abs/2019ApJ...886..102S}
}

@ARTICLE{Shetty_2012,
       author = {{Shetty}, Rahul and {Beaumont}, Christopher N. and {Burton}, Michael G. and {Kelly}, Brandon C. and {Klessen}, Ralf S.},
        title = "{The linewidth-size relationship in the dense interstellar medium of the Central Molecular Zone}",
      journal = {\mnras},
         year = 2012,
        month = sep,
       volume = {425},
       number = {1},
        pages = {720-729},
          doi = {10.1111/j.1365-2966.2012.21588.x},
archivePrefix = {arXiv},
       eprint = {1206.5803},
 primaryClass = {astro-ph.GA},
       adsurl = {https://ui.adsabs.harvard.edu/abs/2012MNRAS.425..720S}
}

@ARTICLE{GRAVITY_2019,
       author = {{GRAVITY Collaboration} and {Abuter}, R. and {Amorim}, A. and {Baub{\"o}ck}, M. and {Berger}, J.~P. and {Bonnet}, H. and {Brandner}, W. and {Cl{\'e}net}, Y. and {Coud{\'e} Du Foresto}, V. and {de Zeeuw}, P.~T. and {Dexter}, J. and {Duvert}, G. and {Eckart}, A. and {Eisenhauer}, F. and {F{\"o}rster Schreiber}, N.~M. and {Garcia}, P. and {Gao}, F. and {Gendron}, E. and {Genzel}, R. and {Gerhard}, O. and {Gillessen}, S. and {Habibi}, M. and {Haubois}, X. and {Henning}, T. and {Hippler}, S. and {Horrobin}, M. and {Jim{\'e}nez-Rosales}, A. and {Jocou}, L. and {Kervella}, P. and {Lacour}, S. and {Lapeyr{\`e}re}, V. and {Le Bouquin}, J. -B. and {L{\'e}na}, P. and {Ott}, T. and {Paumard}, T. and {Perraut}, K. and {Perrin}, G. and {Pfuhl}, O. and {Rabien}, S. and {Rodriguez Coira}, G. and {Rousset}, G. and {Scheithauer}, S. and {Sternberg}, A. and {Straub}, O. and {Straubmeier}, C. and {Sturm}, E. and {Tacconi}, L.~J. and {Vincent}, F. and {von Fellenberg}, S. and {Waisberg}, I. and {Widmann}, F. and {Wieprecht}, E. and {Wiezorrek}, E. and {Woillez}, J. and {Yazici}, S.},
        title = "{A geometric distance measurement to the Galactic center black hole with 0.3\% uncertainty}",
      journal = {\aap},
         year = 2019,
        month = may,
       volume = {625},
          eid = {L10},
        pages = {L10},
          doi = {10.1051/0004-6361/201935656},
archivePrefix = {arXiv},
       eprint = {1904.05721},
 primaryClass = {astro-ph.GA},
       adsurl = {https://ui.adsabs.harvard.edu/abs/2019A&A...625L..10G}
}

@article{Walker_2017,
   title={Star formation in a high-pressure environment: an SMA view of the Galactic Centre dust ridge},
   volume={474},
   ISSN={1365-2966},
   url={http://dx.doi.org/10.1093/mnras/stx2898},
   DOI={10.1093/mnras/stx2898},
   number={2},
   journal={Monthly Notices of the Royal Astronomical Society},
   publisher={Oxford University Press (OUP)},
   author={Walker, D L and Longmore, S N and Zhang, Q and Battersby, C and Keto, E and Kruijssen, J M D and Ginsburg, A and Lu, X and Henshaw, J D and Kauffmann, J and Pillai, T and Mills, E A C and Walsh, A J and Bally, J and Ho, L C and Immer, K and Johnston, K G},
   year={2017},
   month=nov, pages={2373–2388} }

@ARTICLE{Walker_2021,
       author = {{Walker}, Daniel L. and {Longmore}, Steven N. and {Bally}, John and {Ginsburg}, Adam and {Kruijssen}, J.~M. Diederik and {Zhang}, Qizhou and {Henshaw}, Jonathan D. and {Lu}, Xing and {Alves}, Jo{\~a}o and {Barnes}, Ashley T. and {Battersby}, Cara and {Beuther}, Henrik and {Contreras}, Yanett A. and {G{\'o}mez}, Laura and {Ho}, Luis C. and {Jackson}, James M. and {Kauffmann}, Jens and {Mills}, Elisabeth A.~C. and {Pillai}, Thushara},
        title = "{Star formation in 'the Brick': ALMA reveals an active protocluster in the Galactic centre cloud G0.253+0.016}",
      journal = {\mnras},
         year = 2021,
        month = may,
       volume = {503},
       number = {1},
        pages = {77-95},
          doi = {10.1093/mnras/stab415},
archivePrefix = {arXiv},
       eprint = {2102.03560},
 primaryClass = {astro-ph.GA},
       adsurl = {https://ui.adsabs.harvard.edu/abs/2021MNRAS.503...77W}
}

@misc{Walker_2024,
      title={3-D CMZ III: Constraining the 3-D structure of the Central Molecular Zone via molecular line emission and absorption}, 
      author={Daniel L. Walker and Cara Battersby and Dani Lipman and Mattia C. Sormani and Adam Ginsburg and Simon C. O. Glover and Jonathan D. Henshaw and Steven N. Longmore and Ralf S. Klessen and Katharina Immer and Danya Alboslani and John Bally and Ashley Barnes and H Perry Hatchfield and Elisabeth A. C. Mills and Rowan Smith and Robin G. Tress and Qizhou Zhang},
      year={2024},
      eprint={2410.17320},
      archivePrefix={arXiv},
      primaryClass={astro-ph.GA},
      url={https://arxiv.org/abs/2410.17320}, 
}

@ARTICLE{Wang_2014,
       author = {{Wang}, Ke and {Zhang}, Qizhou and {Testi}, Leonardo and {van der Tak}, Floris and {Wu}, Yuefang and {Zhang}, Huawei and {Pillai}, Thushara and {Wyrowski}, Friedrich and {Carey}, Sean and {Ragan}, Sarah E. and {Henning}, Thomas},
        title = "{Hierarchical fragmentation and differential star formation in the Galactic `Snake': infrared dark cloud G11.11-0.12}",
      journal = {\mnras},
         year = 2014,
        month = apr,
       volume = {439},
       number = {4},
        pages = {3275-3293},
          doi = {10.1093/mnras/stu127},
archivePrefix = {arXiv},
       eprint = {1401.4157},
 primaryClass = {astro-ph.GA},
       adsurl = {https://ui.adsabs.harvard.edu/abs/2014MNRAS.439.3275W}
}

@article{Zhang_2021,
   title={HII regions and high-mass starless clump candidates: II. Fragmentation and induced star formation at ~0.025 pc scale: an ALMA continuum study},
   volume={646},
   ISSN={1432-0746},
   url={http://dx.doi.org/10.1051/0004-6361/202038421},
   DOI={10.1051/0004-6361/202038421},
   journal={Astronomy \& Astrophysics},
   publisher={EDP Sciences},
   author={Zhang, S. and Zavagno, A. and López-Sepulcre, A. and Liu, H. and Louvet, F. and Figueira, M. and Russeil, D. and Wu, Y. and Yuan, J. and Pillai, T. G. S.},
   year={2021},
   month=feb, pages={A25} }
\bibliographystyle{aasjournal}

\end{document}